\documentclass[5p,final,times,twocolumn]{elsarticle}

\usepackage{amssymb}

\usepackage[utf8]{inputenc}
\usepackage[cmex10]{amsmath}
\usepackage{amsfonts}
\usepackage{algorithmic}
\usepackage{graphicx}
\usepackage{textcomp}
\usepackage{xcolor}
\usepackage{tikz}
\usepackage{makecell}
\usepackage{footnote}
\usepackage{booktabs} 
\usepackage[tight,footnotesize]{subfigure}
\usepackage{multirow}
\usepackage{threeparttable}
\makesavenoteenv{table}
\makesavenoteenv{tabular}
\usepackage[linewidth=0.5pt]{mdframed}
\usepackage{bbding}

\newcommand\sbullet[1][.66]{\mathbin{\vcenter{\hbox{\scalebox{#1}{$\bullet$~}}}}}

\journal{Journal of Systems and Software}

\begin{document}

\begin{frontmatter}



\title{Using Source Code Density to Improve the Accuracy of Automatic Commit Classification into Maintenance Activities}


\author[lnu]{Sebastian Hönel}
\ead{sebastian.honel@lnu.se}

\author[lnu]{Morgan Ericsson}
\ead{morgan.ericsson@lnu.se}

\author[lnu]{Welf Löwe}
\ead{welf.lowe@lnu.se}

\author[lnu]{Anna Wingkvist}
\ead{anna.wingkvist@lnu.se}

\address[lnu]{Faculty of Technology - Department of Computer Science and Media Technology, Linnaeus University, 351 95 Växjö, Sweden}

\begin{abstract}

Source code is changed for a reason, e.g., to adapt, correct, or adapt it. This reason can provide valuable insight into the development process but is rarely explicitly documented when the change is committed to a source code repository. Automatic commit classification uses features extracted from commits to estimate this reason. 

We introduce \emph{source code density}, a measure of the net size of a commit, and show how it improves the accuracy of automatic commit classification compared to previous size-based classifications. We also investigate how preceding generations of commits affect the class of a commit, and whether taking the code density of previous commits into account can improve the accuracy further.

We achieve up to 89\% accuracy and a Kappa of 0.82 for the cross-project commit classification where the model is trained on one project and applied to other projects. Models trained on single projects yield accuracies of up to 93\% with a Kappa approaching 0.90. The accuracy of the automatic commit classification has a direct impact on software (process) quality analyses that exploit the classification, so our improvements to the accuracy will also improve the confidence in such analyses.
\end{abstract}


\begin{highlights}
\item Approaches for classifying commits based on size differ, contrary to earlier studies
\item Size of a commit is a significant predictor, with net-size being more important
\item Source Code Density is an in-place replacement for more expensive features
\item Existing approaches can be boosted by more than 13\% in accuracy using Density
\item Using the Density of previous generations boosts classification by more than 17\%
\end{highlights}


\end{frontmatter}



\hyphenation{analy-sis}



\section{Introduction}\label{sec:intro}

Every change to the source code of a software system has a purpose, e.g., to correct, perfect, adapt, or extend the system. This purpose can provide valuable insight into the development process, but is rarely documented as part of the change; developers either forget to do so or rely on default classifications, which are often wrong~\cite{hindle2009automatic}. If we could automatically determine the purpose of a change, we could improve the documentation of the change and detect the kind of work done in a software project. This could support, e.g., to identify behavioral patterns, i.e., the developers' behavior and interaction with source code repositories. Such approaches put the developers' work in focus, augment their maintenance profile~\cite{levin2016using}, and influence the development team composition. Regardless of the purpose, it is desirable to move away from error-prone or subjective classification of changes and to introduce objective approaches, as the accuracy to which we can determine the purpose of a change has a direct impact on the validity of our conclusions.

We suggest that \emph{source code density}, the ratio between net- and gross size, can improve the accuracy of change classification. We define net size as the size of the unique code in the system and gross size as the size of everything, including clones, comments, and whitespace. \citet{honel2018poster} studied the size of changes to source code and found significant variances of the source code density but only a weak correlation to the change sizes. The purpose of a change could explain these variances. We measure the density of source code on commit-level. A commit may affect multiple files and different types of changes. We compare the gross-size of that commit, i.e., the sum of files or lines it affects, to its net-size, which is derived by reducing each change to its actual functionality. In subsection~\ref{ssec:size_importance}, we outline how we define source code density, how it can improve the results of automated commit classification, and why commit size is an important predictor.

Previous work by ~\citet{mockus2000identifying} considered features such as meta-information from the committed change, e.g., keywords and comments, properties of the changed source code, and external meta-information from, for example, the bug tracking systems. We hypothesize that changes to source code density better reflect the purpose and that this alone or in combination with previously considered features can improve the classification accuracy. The source code density of a change is, just like the size of the change, cheaper (in terms of effort and computation) to obtain or more convenient to use than some of the other features previously considered. So, even if a combination of features is needed, source code density may be used as a drop-in replacement for some more expensive or inconvenient features. It is noteworthy that the density is a language-agnostic metric that does not require compilation of the underlying software, hence its inexpensiveness. To measure the effectiveness of our proposed features, we reproduce the current state of the art, then add to it, then derive from it, and finally suggest a combined model that delivers the best possible accuracy, improving the state of the art by double digits.

Software re-engineering and maintenance constitute a large part of acquiring knowledge about a system. Large portions of the knowledge in software systems are tacit or inaccessible. While external information and documentation may be used to gather knowledge, those are not always available. It is estimated that up to $60$\% of maintenance work is actually spent on comprehension~\cite{kuhn2007semantic, abran2004software}. Automatic classification of changes has many applications and may help to reduce this time drastically. For instance, it allows us to understand the quality-related aspects of the software development process better. \emph{Software aging} may be avoided by making \emph{change} central in such processes~\cite{fluri2006classifying}. Changes indicate that the process alternates between maintenance phases. In modern projects, features are developed in parallel, bugs are fixed out of band, and maintenance can be done during any of these activities. So, phases can and do overlap, which emphasizes the importance of understanding all ongoing phases.

Such improved understanding can be exploited in a multitude of ways, such as planning resources and personnel for maintenance activities, or to validate that the correct or expected type of planned work is carried out. This is particularly important for projects that are supposedly, e.g., in a feature-freeze phase, as in such phases, no adaptive activities shall be carried out. The type of carried out activity might also be used as a quality indicator when examining the activities' ratio over time, as one could expect, e.g., a project with more perfective and corrective than adaptive commits to be of comparatively higher quality.

We underline two aspects of the software development process in particular that commit classification has a high potential of improving: \emph{process pattern detection} and \emph{software quality monitoring}. A process pattern is an observable and reoccurring sequence of activities~\cite{rising2000scrum,alexander1977pattern} followed in a software development lifecycle. Others have shown that identifying such patterns is valuable~\cite{fluri2008discovering}, as it allows for, e.g., demonstrating that coding guidelines are not followed, or that newcomers lack sufficient training. While some patterns support the process, others are harmful and can be classified as anti-patterns. Various cures to each anti-pattern exist, but it is vital to detect them early to deal with them efficiently. Otherwise, they might result in delays or a decline in productivity or quality. Anti-pattern detection is often governed by data from Application Lifecycle Management (ALM) tools. Such tools extract data from project management applications to draw conclusions from ongoing and historical activities. However, they do not consider the underlying software artifacts that are developed or maintained~\cite{picha2017towards}. Classifying the current activities can reduce ambiguities in detecting patterns by their symptoms. Others have demonstrated that such pattern detection best involves project-level metrics as well as developer-level information~\cite{levin2016using}. A short anecdote may emphasize our case: During a collaboration, where our colleagues had access to such ALM data, we brought in quality information about the source code and were able to detect process anti-patterns, such as \emph{Nine Pregnant Women}\footnote{https://github.com/ReliSA/Software-process-antipatterns-catalogue/blob/master/catalogue/Nine\textunderscore Pregnant\textunderscore Women.md}, or the \emph{Lone Wolf Programmer}\footnote{https://github.com/ReliSA/Software-process-antipatterns-catalogue/blob/master/catalogue/Lone-Wolf.md}. However, since some patterns share certain symptoms, we were unable to distinguish which pattern occurred in certain phases. Our situation would have been relieved by having a proper commit classification at our disposal. This was the initial incentive to begin work on this study and to incorporate data and tools we already had at our disposal.

The central question of this research is whether the source code density can improve the accuracy of classifying a change by its purpose. We rely on the definitions of maintenance activities by \citet{mockus2000identifying}, and classify changes as either \emph{(a)daptive}, \emph{(c)orrective}, or \emph{(p)erfective}. Adaptive activities add new features, corrective activities fix faults, and perfective activities restructure code to accommodate future changes.


The paper is structured as follows. Section~\ref{sec:background} provides a deep and qualitative insight into the importance of size and source code density. Section~\ref{sec:method} introduces the research questions and the approach to answer them. Section~\ref{sec:result} presents the results. Section~\ref{sec:threads} discusses threats to validity. Section~\ref{sec:rel_work} gives an overview of related work. Section~\ref{sec:conclusions} concludes the research and section ~\ref{sec:future} shows directions of future work.

\begin{figure}
    \centering
\begin{tikzpicture}[x=1pt,y=1pt]
\definecolor{fillColor}{RGB}{255,255,255}
\path[use as bounding box,fill=fillColor,fill opacity=0.00] (0,0) rectangle (251.50,158.99);
\begin{scope}
\path[clip] (  0.00,  0.00) rectangle (251.50,158.99);
\definecolor{drawColor}{RGB}{255,255,255}
\definecolor{fillColor}{RGB}{255,255,255}

\path[draw=drawColor,line width= 0.5pt,line join=round,line cap=round,fill=fillColor] (  0.00,  0.00) rectangle (251.50,158.99);
\end{scope}
\begin{scope}
\path[clip] ( 42.30,  6.75) rectangle (154.56,154.49);
\definecolor{fillColor}{RGB}{255,255,255}

\path[fill=fillColor] ( 42.30,  6.75) rectangle (154.56,154.49);
\definecolor{drawColor}{gray}{0.87}

\path[draw=drawColor,line width= 0.1pt,line join=round] ( 42.30, 29.85) --
	(154.56, 29.85);

\path[draw=drawColor,line width= 0.1pt,line join=round] ( 42.30, 62.60) --
	(154.56, 62.60);

\path[draw=drawColor,line width= 0.1pt,line join=round] ( 42.30, 95.36) --
	(154.56, 95.36);

\path[draw=drawColor,line width= 0.1pt,line join=round] ( 42.30,128.12) --
	(154.56,128.12);

\path[draw=drawColor,line width= 0.2pt,line join=round] ( 42.30, 13.47) --
	(154.56, 13.47);

\path[draw=drawColor,line width= 0.2pt,line join=round] ( 42.30, 46.22) --
	(154.56, 46.22);

\path[draw=drawColor,line width= 0.2pt,line join=round] ( 42.30, 78.98) --
	(154.56, 78.98);

\path[draw=drawColor,line width= 0.2pt,line join=round] ( 42.30,111.74) --
	(154.56,111.74);

\path[draw=drawColor,line width= 0.2pt,line join=round] ( 42.30,144.50) --
	(154.56,144.50);

\path[draw=drawColor,line width= 0.2pt,line join=round] ( 72.91,  6.75) --
	( 72.91,154.49);

\path[draw=drawColor,line width= 0.2pt,line join=round] (123.94,  6.75) --
	(123.94,154.49);
\definecolor{fillColor}{RGB}{227,26,28}

\path[fill=fillColor] ( 49.95, 13.47) rectangle ( 95.88, 13.47);
\definecolor{fillColor}{RGB}{251,154,153}

\path[fill=fillColor] ( 49.95, 13.47) rectangle ( 95.88, 13.47);
\definecolor{fillColor}{RGB}{51,160,44}

\path[fill=fillColor] ( 49.95, 13.47) rectangle ( 95.88, 13.47);
\definecolor{fillColor}{RGB}{178,223,138}

\path[fill=fillColor] ( 49.95, 13.47) rectangle ( 95.88, 13.47);
\definecolor{fillColor}{RGB}{31,120,180}

\path[fill=fillColor] ( 49.95, 13.47) rectangle ( 95.88, 13.47);
\definecolor{fillColor}{RGB}{166,206,227}

\path[fill=fillColor] ( 49.95, 13.47) rectangle ( 95.88,147.78);
\definecolor{fillColor}{RGB}{227,26,28}

\path[fill=fillColor] (100.98, 13.47) rectangle (146.90, 36.40);
\definecolor{fillColor}{RGB}{251,154,153}

\path[fill=fillColor] (100.98, 36.40) rectangle (146.90, 72.43);
\definecolor{fillColor}{RGB}{51,160,44}

\path[fill=fillColor] (100.98, 72.43) rectangle (146.90, 85.54);
\definecolor{fillColor}{RGB}{178,223,138}

\path[fill=fillColor] (100.98, 85.54) rectangle (146.90,115.02);
\definecolor{fillColor}{RGB}{31,120,180}

\path[fill=fillColor] (100.98,115.02) rectangle (146.90,147.78);
\definecolor{fillColor}{RGB}{166,206,227}

\path[fill=fillColor] (100.98,147.78) rectangle (146.90,147.78);
\definecolor{drawColor}{gray}{0.70}

\path[draw=drawColor,line width= 0.5pt,line join=round,line cap=round] ( 42.30,  6.75) rectangle (154.56,154.49);
\end{scope}
\begin{scope}
\path[clip] (  0.00,  0.00) rectangle (251.50,158.99);
\definecolor{drawColor}{gray}{0.30}

\node[text=drawColor,anchor=base east,inner sep=0pt, outer sep=0pt, scale=  0.72] at ( 38.25, 10.99) {0};

\node[text=drawColor,anchor=base east,inner sep=0pt, outer sep=0pt, scale=  0.72] at ( 38.25, 43.75) {50};

\node[text=drawColor,anchor=base east,inner sep=0pt, outer sep=0pt, scale=  0.72] at ( 38.25, 76.50) {100};

\node[text=drawColor,anchor=base east,inner sep=0pt, outer sep=0pt, scale=  0.72] at ( 38.25,109.26) {150};

\node[text=drawColor,anchor=base east,inner sep=0pt, outer sep=0pt, scale=  0.72] at ( 38.25,142.02) {200};
\end{scope}
\begin{scope}
\path[clip] (  0.00,  0.00) rectangle (251.50,158.99);
\definecolor{drawColor}{gray}{0.70}

\path[draw=drawColor,line width= 0.2pt,line join=round] ( 40.05, 13.47) --
	( 42.30, 13.47);

\path[draw=drawColor,line width= 0.2pt,line join=round] ( 40.05, 46.22) --
	( 42.30, 46.22);

\path[draw=drawColor,line width= 0.2pt,line join=round] ( 40.05, 78.98) --
	( 42.30, 78.98);

\path[draw=drawColor,line width= 0.2pt,line join=round] ( 40.05,111.74) --
	( 42.30,111.74);

\path[draw=drawColor,line width= 0.2pt,line join=round] ( 40.05,144.50) --
	( 42.30,144.50);
\end{scope}
\begin{scope}
\path[clip] (  0.00,  0.00) rectangle (251.50,158.99);
\definecolor{drawColor}{gray}{0.70}

\path[draw=drawColor,line width= 0.2pt,line join=round] ( 72.91,  4.50) --
	( 72.91,  6.75);

\path[draw=drawColor,line width= 0.2pt,line join=round] (123.94,  4.50) --
	(123.94,  6.75);
\end{scope}
\begin{scope}
\path[clip] (  0.00,  0.00) rectangle (251.50,158.99);
\definecolor{drawColor}{RGB}{0,0,0}

\node[text=drawColor,rotate= 90.00,anchor=base,inner sep=0pt, outer sep=0pt, scale=  0.90] at ( 10.70, 80.62) {Lines of Code};
\end{scope}
\begin{scope}
\path[clip] (  0.00,  0.00) rectangle (251.50,158.99);
\definecolor{fillColor}{RGB}{255,255,255}

\path[fill=fillColor] (163.56, 21.54) rectangle (247.00,139.71);
\end{scope}
\begin{scope}
\path[clip] (  0.00,  0.00) rectangle (251.50,158.99);
\definecolor{drawColor}{RGB}{0,0,0}

\node[text=drawColor,anchor=base west,inner sep=0pt, outer sep=0pt, scale=  0.90] at (168.06,128.13) {Type};
\end{scope}
\begin{scope}
\path[clip] (  0.00,  0.00) rectangle (251.50,158.99);
\definecolor{fillColor}{RGB}{255,255,255}

\path[fill=fillColor] (168.06, 98.31) rectangle (182.51,112.76);
\end{scope}
\begin{scope}
\path[clip] (  0.00,  0.00) rectangle (251.50,158.99);
\definecolor{fillColor}{RGB}{166,206,227}

\path[fill=fillColor] (168.77, 99.02) rectangle (181.80,112.05);
\end{scope}
\begin{scope}
\path[clip] (  0.00,  0.00) rectangle (251.50,158.99);
\definecolor{fillColor}{RGB}{255,255,255}

\path[fill=fillColor] (168.06, 83.85) rectangle (182.51, 98.31);
\end{scope}
\begin{scope}
\path[clip] (  0.00,  0.00) rectangle (251.50,158.99);
\definecolor{fillColor}{RGB}{31,120,180}

\path[fill=fillColor] (168.77, 84.56) rectangle (181.80, 97.59);
\end{scope}
\begin{scope}
\path[clip] (  0.00,  0.00) rectangle (251.50,158.99);
\definecolor{fillColor}{RGB}{255,255,255}

\path[fill=fillColor] (168.06, 69.40) rectangle (182.51, 83.85);
\end{scope}
\begin{scope}
\path[clip] (  0.00,  0.00) rectangle (251.50,158.99);
\definecolor{fillColor}{RGB}{178,223,138}

\path[fill=fillColor] (168.77, 70.11) rectangle (181.80, 83.14);
\end{scope}
\begin{scope}
\path[clip] (  0.00,  0.00) rectangle (251.50,158.99);
\definecolor{fillColor}{RGB}{255,255,255}

\path[fill=fillColor] (168.06, 54.94) rectangle (182.51, 69.40);
\end{scope}
\begin{scope}
\path[clip] (  0.00,  0.00) rectangle (251.50,158.99);
\definecolor{fillColor}{RGB}{51,160,44}

\path[fill=fillColor] (168.77, 55.66) rectangle (181.80, 68.69);
\end{scope}
\begin{scope}
\path[clip] (  0.00,  0.00) rectangle (251.50,158.99);
\definecolor{fillColor}{RGB}{255,255,255}

\path[fill=fillColor] (168.06, 40.49) rectangle (182.51, 54.94);
\end{scope}
\begin{scope}
\path[clip] (  0.00,  0.00) rectangle (251.50,158.99);
\definecolor{fillColor}{RGB}{251,154,153}

\path[fill=fillColor] (168.77, 41.20) rectangle (181.80, 54.23);
\end{scope}
\begin{scope}
\path[clip] (  0.00,  0.00) rectangle (251.50,158.99);
\definecolor{fillColor}{RGB}{255,255,255}

\path[fill=fillColor] (168.06, 26.04) rectangle (182.51, 40.49);
\end{scope}
\begin{scope}
\path[clip] (  0.00,  0.00) rectangle (251.50,158.99);
\definecolor{fillColor}{RGB}{227,26,28}

\path[fill=fillColor] (168.77, 26.75) rectangle (181.80, 39.78);
\end{scope}
\begin{scope}
\path[clip] (  0.00,  0.00) rectangle (251.50,158.99);
\definecolor{drawColor}{RGB}{0,0,0}

\node[text=drawColor,anchor=base west,inner sep=0pt, outer sep=0pt, scale=  0.72] at (187.01,103.05) {Gross size};
\end{scope}
\begin{scope}
\path[clip] (  0.00,  0.00) rectangle (251.50,158.99);
\definecolor{drawColor}{RGB}{0,0,0}

\node[text=drawColor,anchor=base west,inner sep=0pt, outer sep=0pt, scale=  0.72] at (187.01, 88.60) {Net-Functionality};
\end{scope}
\begin{scope}
\path[clip] (  0.00,  0.00) rectangle (251.50,158.99);
\definecolor{drawColor}{RGB}{0,0,0}

\node[text=drawColor,anchor=base west,inner sep=0pt, outer sep=0pt, scale=  0.72] at (187.01, 74.15) {Clones};
\end{scope}
\begin{scope}
\path[clip] (  0.00,  0.00) rectangle (251.50,158.99);
\definecolor{drawColor}{RGB}{0,0,0}

\node[text=drawColor,anchor=base west,inner sep=0pt, outer sep=0pt, scale=  0.72] at (187.01, 59.69) {Dead code};
\end{scope}
\begin{scope}
\path[clip] (  0.00,  0.00) rectangle (251.50,158.99);
\definecolor{drawColor}{RGB}{0,0,0}

\node[text=drawColor,anchor=base west,inner sep=0pt, outer sep=0pt, scale=  0.72] at (187.01, 45.24) {Whitespace};
\end{scope}
\begin{scope}
\path[clip] (  0.00,  0.00) rectangle (251.50,158.99);
\definecolor{drawColor}{RGB}{0,0,0}

\node[text=drawColor,anchor=base west,inner sep=0pt, outer sep=0pt, scale=  0.72] at (187.01, 30.78) {Comments};
\end{scope}
\end{tikzpicture}
    \vspace{-10pt}
    \caption{Exemplary comparison of sizes measured for a single file containing some typical source code.}
    \label{fig:density_barplots}
\end{figure}
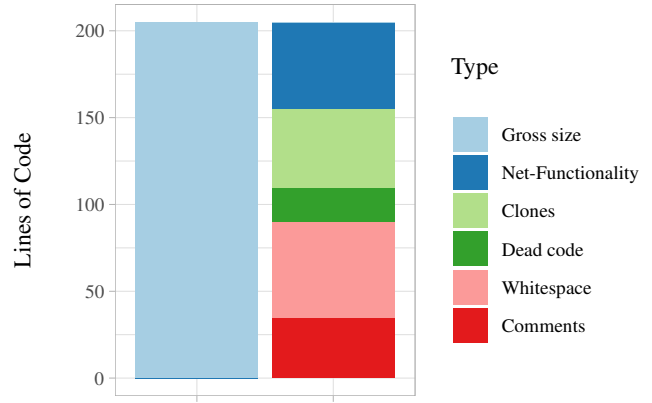

\section{Background}\label{sec:background}

We have previously examined the impact of \emph{code density} on effort- and productivity estimations \cite{honel2018poster}. Due to the unavailability of precise measures of spent time, we were unable to establish strong correlations. However, we found significant deviations of code density and size for various notions of the size of code committed to software repositories.

In this section, we first elaborate on the importance of change size and the potential of density. Then, we present the most relevant studies for size-based applications. The section is concluded by introducing the extended dataset used throughout this study, and size-based metrics therein. 

\subsection{The Importance of Size and the Potential of Density} \label{ssec:size_importance}

There are various ways of quantifying the number of changes in (or the size of) a commit, as outlined in section~\ref{sec:rel_work}. This study focuses on measuring the size using varying forms of lines of code (LOC), however.
Maintenance of software is such an integral part of its evolutionary process that it consumes much of the total resources available, according to a field study carried out by~\citet{lientz1978characteristics}. At the same time, maintenance phases lack a sufficiently strong understanding.

In its simplest form, determining the size by counting LOC neglects what these lines comprise. A typical example of a file containing source code is given in Figure~\ref{fig:density_barplots}. In it, we usually find \emph{cloned code}, that is, functional equivalent or even identical code that is to be found in at least one other file, or even in another portion of the same file. Dead code is an aggregation of code that cannot be called, such as statements that occur after the return statement (where valid) or non-reachable \emph{if}/\emph{else}-branches, and code that is never called within the application. Whitespace is any excessive empty lines and empty characters that do not contribute to the code's functionality. Comments, while useful or even necessary, are not counted, either. This is because of how we define the Density:
\begin{align}
    \textit{Density} &= \frac{\text{Functionality}}{\text{Size\textsubscript{gross}}}.
\end{align}
The functionality comprises all code that purely contributes to the application's functioning; thus it does not include comments, whitespace, cloned (duplicated) functionality, and dead code. Hence, the density can maximally approach $1$, and minimally be $0$.

Determining the density for a single file is useful, and extending the approach to the entirety of a software may have a significant meaning for estimation models. For example, the International Software Benchmarking Standards Group (ISBSG) further defines the project metrics \emph{effort} and \emph{productivity}, and bases them mainly on the size of software\footnote{ISBSG: ``Software size as the main input parameter to cost estimation models.'', http://isbsg.org/software-size}, and change over time thereof:
\begin{align}
        \textit{Effort\textsubscript{ISBSG}} &= \frac{\text{Size\textsubscript{gross}}}{\text{time}} \label{eq:eff_isbsg} \\[10pt]
        \textit{Effort\textsubscript{net}} &= \frac{\text{Density} \times \text{Size\textsubscript{gross}}}{\text{time}} \label{eq:eff_dens} \\[10pt]
                                           &= \frac{\text{Functionality}}{\text{time}}          \label{eq:eff_dens2}
\end{align}
Equations~\eqref{eq:eff_isbsg} and~\eqref{eq:prod_isbsg} show how ISBSG defines effort (size per time) and productivity (effort per size). The most common method used by them for determining size is function points (FP). A function point is a unit of measurement to express the amount of business functionality in a software. Various types of FP exist and those are specified mostly as ISO standards. FP are usually counted manually. \citeauthor{albrecht1983software} found a strong positive correlation between FP and LOC~\cite{albrecht1983software}, thus questioning the value of FP, given its cost compared to counting LOC. Adapting the ISBSG-equations to use notions of net-size ($\textit{Size\textsubscript{net}} = \text{Density} * \text{Size\textsubscript{gross}}$) instead of gross-size or size measured in terms of function points, leads to equations~\eqref{eq:eff_dens},~\eqref{eq:eff_dens2} and~\eqref{eq:prod_dens}.
\begin{align}
    \textit{Productivity\textsubscript{ISBSG}} &= \frac{\text{Effort}}{\text{Size\textsubscript{gross}}} \label{eq:prod_isbsg}   \\[10pt]
    \textit{Productivity\textsubscript{net}} &= \frac{\text{Density}}{\text{time}}          \label{eq:prod_dens}
\end{align}
Estimation models with a strong focus on software size may behave differently, given these alterations. In this study, we report significant deviations between the net- and gross-size of software. While strongly positively correlated, the correlation is non-linear. Estimating the net-size, if counted as LOC, can be automated conveniently, unlike counting function points. Furthermore, while net-size allows a better approximation of effort and productivity, it may also allow increasing the confidence in automatic commit classification. This comes into play when the different activities in a software project are estimated individually.

\subsection{The Most Relevant Studies for Size and Density} \label{ssec:bg_extra_relevant}
It is crucial to outline why our study, which evolves around the size of commits, is important. More specifically, the following short qualitative study of the most relevant related work emphasizes the two questions:

\begin{mdframed}[nobreak=true]
    $\sbullet$~Is the size of a commit an important predictor? \\
    $\sbullet$~Why may \emph{source code density} improve the results of automated commit classification?
\end{mdframed}

The size of software, regardless of how it is measured, is often considered a low-level metric for software and its evolutionary process~\cite{herraiz2006comparison}. While it is a simple and thus computationally cheap metric, compared to metrics such as cyclomatic complexity or coupling/cohesion, discourse about its applicability and how it should be obtained, exist.~\citeauthor{herraiz2006comparison} point out that discrepancies about measurements of size, especially between \emph{libre}  (free, open source) and \emph{'traditional'} software, exist. While the method of measuring the size is different, the evolution of software belonging to either system, according to the laws of software evolution by~\citet{lehman1985program}, appears to be the same. This work is only concerned with libre software. We report significant deviations of the obtained net- and gross size measurements. In the context of software evolution analysis, the automated classification of commits, based on density rather than on raw LOC, has the potential to resemble the actual size of software more closely, and thus to improve the analysis process. A repetition of the study by~\citet{herraiz2006comparison} using density is thus likely to yield different results.

The size of a commit can also reveal developmental aspects and practices of the software evolutionary process, as studied by~\citet{hindle2008large}. Large commits, for example, often happen when branches in a repository are merged, which hints at a process where features and bugfixes are developed separately and then integrated once they achieve a certain level of maturity or tests are passed. It has become quite common to rely on external packages that are downloaded whenever the software is built or run for the first time. However, there is also software that incorporates external code, such as libraries or frameworks, instead of referencing it~\cite{hindle2008large}. Large commits reflect such behavior in the first place  and can give an indication of the software's maintainability, as incorporated code is much less frequently updated. Large commits may also be more correlated with automatically generated code or documentation. As for the relation to this work, it was found that large commits are most often \emph{perfective}. The size is, therefore, relevant to reveal such aspects. Using the density instead of (or in conjunction with) the size may help to better identify developmental aspects and to reduce ambiguities that would otherwise arise from only counting LOC. Like the metadata of commits, such as its keywords and message, author, and timestamps etc., is attractive because it does not require any extra computation. The same is true for LOC-based size methods, which are, on top of that, language-agnostic. Such methods can, therefore, be effortlessly integrated into existing metadata-based solutions and improve and aid the automatic classification, which is still a challenge~\cite{hindle2008large}. Having an automated classifier is desirable, especially for software where tagging the commits with a purpose was not previously enforced or must be done retroactively.

Small commits have shown to be a significant predictor for faults~\cite{leszak2002classification}, with LOC being an effective predictor~\cite{bell2011does}. \citeauthor{purushothaman2005toward} point out that the impact of small changes to source code is often underrated~\cite{purushothaman2005toward}. This leads to less rigorous processes in the software evolutionary process, such as lack of testing. The implications on the system's architecture originating from small changes are often associated with small risk, too. There are cases where one-line changes cost more than one billion US dollars, as reported by~\citet{weinberg1983kill}. As we demonstrate, the majority of such small changes are related to correcting bugs, followed by introducing new features. These are also the activities mainly correlated with introducing faults. An automated commit classification could aid the process of flagging such potentially fault-introducing commits for a more thorough audit. Whether small, large, or anywhere in between, the size of a commit is a piece of valuable information worth exploiting. \citet{mockus2000identifying} report strong relationships between the type and size of a change. The impact of using density instead of raw LOC is significant when classifying commits, and hence to triage them for further inspection.

\subsection{The Extended Dataset}\label{ssec_dataset}

Our extended dataset~\cite{honel2019commits} is based on the dataset by \citet{levin2017boosting} but extended with size data. The original dataset~\cite{levin2017commits} consists of more than 1\,150 manually labeled commits from eleven projects. We used our tool suite, \emph{\mbox{Git-Density}}~\cite{honel2019gitdensity}, to collect size data for the 11 projects. For all but the two projects, namely \emph{Intellij Community Edition} and \emph{Kotlin}, we have added size data for all commits of each project's repository, as of January 2019. From those two projects, we have analyzed all commits that were contained in the initial dataset, plus the first 30\,000 and 35\,000 commits respectively. We merged the two datasets using each commit's unique SHA1 hash. During this process, we identified two duplicate commits in the original dataset, effectively reducing it to 1\,149 samples.



We define the \emph{size} of a commit to be either the number of files or the LOC that were changed, across all types of changes, i.e., files/lines added, deleted, modified, or renamed. The \emph{Gross size} is the size without considering whether a line affects the functionality of the source code or not. \emph{Net size}, on the other hand, is the gross size minus the number of files or LOC that did not contribute to actual changes to the software's functionality. We consider empty lines, whitespace, as well as single- and multi-line comments to be without such effect. Conversely, if any other source code line was changed, it is undecidable whether or not the functionality changes, and we conservatively assume it does. If none of the changed lines in a file is considered to contribute to changed functionality, then neither is the file.


The \emph{Change density} (or short density) is the ratio between \emph{net} to \emph{gross} size of a change. If all lines changed in a commit potentially contribute to the software's functionality, then the density takes its maximum value of $1.0$. Conversely, if no line changes, the density takes its lowest value of $0.0$.

The extended dataset contains the following eight features (gross and net sizes) that describe the number of files that have been added, deleted, renamed, or modified in a commit. Renaming a file means that a file is deleted in one place and reappears in another place, without having its content changed (pure rename). If its content is similar by $50$\% or more (but less than $100$\%), the change is considered an impure rename (common git threshold). If the similarity undercuts the threshold, the commit exhibits one deleted and one added file instead. For brevity, we sometimes refer to a feature by its number (or its number followed by an \textbf{a} if we refer to its corresponding net-version).


\renewcommand{\theenumi}{\arabic{enumi}}
\begin{enumerate}
    \setcounter{enumi}{0}
    \item Number of Files Added (Gross and Net)
    \item Number of Files Deleted (Gross and Net)
    \item Number of Files Renamed (Gross and Net)
    \item Number of Files Modified (Gross and Net)
\end{enumerate}

There may be one or more changes per file. These are called \emph{Hunks}. We can determine the density of an entire file in a commit by aggregating the properties of its hunks. If the aggregated changes of all hunks of a file amount to zero lines affected, then the respective net-feature does not count the file as being affected. This is an important measure, as previous researchers have also determined the size of a commit by the number of files it affects~\cite{herraiz2006comparison}.

\begin{enumerate}
    \setcounter{enumi}{4}
    \item Number of Lines Added by Added Files (Gross and Net)
    \item Number of Lines Deleted by Deleted Files (Gross and Net)
    \item Number of Lines Added by Modified Files (Gross and Net)
    \item Number of Lines Deleted by Modified Files (Gross and Net)
    \item Number of Lines Added by Renamed Files (Gross and Net)
    \item Number of Lines Deleted by Renamed Files (Gross and Net)
    \item Affected Files Ratio Net
    \item Density
\end{enumerate}

An additional feature, Affected Files Ratio Net (11), expresses the ratio between the sums of all gross (1--4) and net (1a--4a) files above. We also added the feature Density (12) to the dataset. It describes the ratio between the two sums of all lines added, deleted, modified, and renamed and their resp. gross-version. A density of zero means that the sum of net-lines is zero (i.e., all lines changed were just clones, dead code, whitespace, comments, etc.). A density of $1.0$ means that all changed net-lines contribute to the gross-size of the commit (i.e., no lines considered useless with, e.g., only comments or whitespace).

These twelve attributes count the gross amount of lines of code affected by added, deleted, modified, and renamed files, while their corresponding net-version counts the net-amount. Note that added files can never go along with deleted lines, and that deleted files can never include added lines. As for renamed files, the standard $50$\% similarity threshold applies; therefore, those can have either type of change.



\section{Methodology}\label{sec:method}

Classification is a common problem of statistics and machine learning. Classification models are fit using labeled data, with the intention to correctly label previously unseen observations based on some of their features. Commit classification models may facilitate and learn from a number of different features to achieve this task of generalizing from examples. Previously, some of these models were based on, e.g., keywords and comments (commit messages)~\cite{mockus2000identifying, levin2017boosting}. Some other models used notions of commit size~\cite{hindle2008large, hindle2009automatic, herraiz2006comparison, purushothaman2005toward}. In this study, we attempt to use a more well-defined size metric, the \emph{density}, to build classifiers that can assign commits to maintenance activities. Such classifiers, once trained, can be used for \emph{automatic} classification of previously unseen and uncategorized commits. Without these, one would have to resort to manually labeling instances, which is error-prone and may require an extensive set of rules.

In the remainder of this section we pose our research questions and devise several experiments to resolve them empirically. We then outline the statistical methods that we apply to examine the gathered data.

\subsection{Research Questions}

\renewcommand{\theenumi}{\Alph{enumi}} 
\vspace{1em}
\textbf{RQ 1:} Does the net-size of commits as gathered in the extended dataset reveal significant differences, compared to the gross-size that was used in prior studies?
\begin{enumerate}
    \item Are the evolutionary patterns the same for classifying commits when gross- and net size of file- and line counts are considered?
    \item Do the size and frequency change when considering the net size?
\end{enumerate}



\textbf{RQ 2:} Using the existing maintenance activities (labels), how well do source code density (including gross- and net size) alone allow for a classification of maintenance activities?
\begin{enumerate}
    \item Is there a difference in accuracy for cross- and single-project classification?
    \item Do the net size features perform better in classification, compared to their gross size counterparts?
\end{enumerate}

\textbf{RQ 3:} Are the size- and density features suitable for improving state of the art in commit classification accuracy?
\begin{enumerate}
    \item Are the previous results as obtained by \citet{levin2017boosting} reproducible?
    \item If we extend their models with size and density data, does the accuracy improve?
    \item Is there a best subset of features, that combines source code density features and those from \citeauthor{levin2017commits}, i.e., a set of best features across all datasets?
\end{enumerate}

\textbf{RQ 4:} What is the impact of size features of commits in previous generations when classifying a (principal) commit?
\begin{enumerate}
    \item What are the most important features of the principal commit?
    \item How important are density- and size features in preceding generations?
    \item Is there a significant difference between cross- and single-project classification?
\end{enumerate}

\subsection{Statistical Methods}

Previous studies~\cite{fernandez2014we} found Random forests to perform well in general. Based on this, we mainly use Random forests to obtain rankings of predictors (variable importance)~\cite{breiman2001random,renv2017}. The research we relate our work to reports classification results for single projects and cross-project. They achieved the best results using Gradient Boosting Machines (GBM)~\cite{friedman2002stochastic} and Random forests. To evaluate and compare their accuracy, we apply the Zero Rule (ZeroR) classifier to set a baseline. For the prediction of categorical variables, that classifier always predicts the most common class. We use \emph{R}~\cite{renv2017} to perform all experiments and analyses.

Throughout this paper we do only report classification accuracy and \citeauthor{cohen1960coefficient}'s Kappa. Some of the related work we refer to reports other or additional metrics, such as precision, recall, or F1-score. However, accuracy and Kappa is to be found in most of the other studies, and thus makes our work comparable to them. Kappa is a metric that, if it is reported along with accuracy, mitigates some of the caveats of the F1-score. It is defined as:
\begin{align}
    \textit{Kappa} &= \frac{\text{Accuracy\textsubscript{total}} - \text{Accuracy\textsubscript{random}}}{1 - \text{Accuracy\textsubscript{random}}}.
\end{align}
Given an uneven or strongly skewed distribution of classes in a dataset, the reported accuracy and F1-score may very well be high using, e.g., the ZeroR classifier, as those metrics do not correct for the bias of such skewed distributions, or how the agreement between raters could occur by chance. \citeauthor{cohen1960coefficient}'s Kappa however corrects for those, and would give an indication of the low agreement between predicted and true classes. We deem the combination of classification accuracy and Kappa as metrics therefore to be sufficient.

Research Question 1 addresses statistical properties, such as distributions of the commits' labels of the extended dataset. These properties are useful for putting the dataset into relation to the work of other researchers, such as \citet{hattori2008nature} and \citet{purushothaman2005toward}. As statistical tools, we make use of (Empirical) Cumulative Distribution Functions (E)CDF and empirical densities to find similarities and differences between the nature of commits concerning the different notions of size. It is important to note that related work used different manually labeled datasets, not a common benchmark suite.

\label{ssec_stat_meth_rq2a}
For Research Question 2A, we apply the methods to the entire dataset and report the models' accuracy. The goal was to understand the importance of each new attribute, not to predict validation samples. Therefore, we apply the following methods across all projects and then to every single project separately:

\begin{itemize}
    \item Remove zero-variance predictors. Within the scope of a single project, some features do not exhibit any variance any longer and are therefore removed. Between projects, such zero-variance features varied.
    \item Identify highly correlated (coefficient larger than $0.75$) features using the \emph{Pearson} co-variance. However, we keep the features for further analyses and eliminate later, when assessing variable importance and doing a separate Recursive Feature Elimination (RFE, elaborated below).
    \item Assess \emph{variable importance} using a Receiver Operating Characteristic (ROC)~\cite{hanley1982meaning} curve analysis by applying the Learning Vector Quantization (LVQ)~\cite{kohonen1995learning} method. We prefer LVQ over other methods, as it reports importance for each label and outperforms other methods. We also tried to use GBMs and \emph{eXtreme Gradient Boosting} (xGB)~\cite{chen2015xgboost}. However, those turned out to be significantly slower (runtime) and do not report importance per label.
    \item Run an RFE across all projects and for each project individually. Attempt to use between just one and all available features (i.e., those that withstood the previous RFE). Use Random forests to extract variable importances, using ten-fold cross-validation, while computing at least three sets of complete folds.
\end{itemize}

The R package \emph{Caret}~\cite{JSSv028i05} implements RFE and provides sets of interchangeable methods to fit models and allows re-sampling using, e.g., cross-validation. In RFE, the underlying fitting method first fits a model to the training data using all of its predictors. Then, low-weight features are removed recursively with each iteration. Ideally, such method also provides the variable importance to rank the features. Random forests hence is a suitable candidate, and we have used it, also because of its favorable accuracy.

For a set $S_i$ of attribute sizes referring to the top-ranked $i$ attributes, the model is refit using those attributes, then that model's accuracy is assessed, and in the end, the best model is retained.

We follow suggestions that recommend having the model selection process use external validation through re-sampling by cross-validation~\cite{ambroise2002selection,svetnik2004application}. The described procedure for model-fitting on best-ranking attribute subsets was therefore nested in a $k$-fold cross-validation, using three or more complete sets of folds.

For Research Question 2B, we compare the net- vs. gross-size attributes of the extended datasets. We report the results for individual projects and across projects. The steps undertaken are:

\label{ssec_stat_meth_rq2b}
\begin{itemize}
    \item Vertically separate the extended dataset into one that contains only the net- and one that contains only the gross-version of each attribute. Both datasets retain the labels.
    \item Assess the variable importance of either dataset using a ROC curve analysis, based on Random forests, using a 10-fold and three times repeated cross-validation.
    \item Repeat the last step and gather results for each project.
\end{itemize}

Research Question 3 is three-fold. Previously, \citet{levin2017boosting} demonstrated strong classification results using commit keywords and source code changes. We are interested in examining whether the predictive power of their models can be further enhanced using size attributes.

\label{ssec_stat_meth_rq3}
\begin{itemize}
    \item Attempt to reproduce the previous researchers' results by using their dataset~\cite{levin2017commits} and methods. They report training- and validation accuracy of J48-, GBM- and Random forest-based models. As their champion model uses Random forests, we only reproduce those models.
    \item Preliminary split the data into $85$\% training and $15$\% validation samples. Then further vertically split the training data into one dataset containing only keywords, one containing only changes and one that combines both of these, so that we may reconstruct all types of classifiers used.
    \item Use the custom classifier they suggest for compound models (see below this list). The combinations of two models $A, B$ in shape of $\{A,B\}$ and $\{B,A\}$ are considered to be distinct by that classifier.
    \item Build and train three different models (one per subdivided dataset), using Random forests and five times repeated 10-fold cross-validation (that validation happens entirely on the $85$\% training data).
    \item Construct the compound models. Compound models have a left and a right model. For those models based on just one type of model (e.g., keywords), use the same model on both sides.
    \item Run the custom classifier on each compound model. The classifier uses the left model whenever a sample uses at least one keyword out of the 20.
    \item Run the classifier on the $15$\% of the previously unseen validation samples, report accuracy during training and validation, and compare. We combine the numeric votes for each class of each model and select the highest (the models return a probability for each class) when we report the training accuracy for compound models.
\end{itemize}

There are a total of nine compound models, as there are three types of possible underlying models $\{\mathit{keywords}, \mathit{changes}, \mathit{combined}\}$. The nine models are the result of building all permutations ($3^2$). A compound model is the combination of two models (a ``classifier lattice'', cf. ~\citet{levin2017boosting}), such that the routine for classifying a commit uses a different model, depending on whether the commit's message has any of the keywords the keyword-classifier was trained on. This notion of a 2-compound model was introduced, as the keyword-based classifiers outperformed the other classifiers if keywords were present. These compound models do not overlap because each single model may or may not be a reduced-feature model.

For simplicity, we refer to the two models in a compound model simply as left and right model. For the second part of this research question, we add one more type of underlying model, namely $model\textsubscript{density}$, resulting in 16 compound models ($4^2$). That model is based and trained on size data only. Also, we alter the $model\textsubscript{combined}$ model to also span the size attributes. The models using $\{\mathit{keywords}, \mathit{changes}\}$ remain the same. We then apply the same procedures as in the previous list to report training- and validation-accuracy.


As for the last part of Research Question 3, we attempt to further tune and prune the 16 compound models. The steps involved are:

\begin{itemize}
    \item Create a density-only dataset, based only on net-attributes.
    \item Attempt further optimizations to that dataset, by conditionally leaving out zero- and near-zero-variance attributes and preprocessing it. Attempt various (combinations of) preprocessing, such as scaling (divide by mean), centering (subtract mean), or Yeo–Johnson transformations~\cite{yeo2000new} (suitable as we are dealing with power-distributed data that can be zero).
    \item Analyze the variable importance of that dataset to find the optimum amount of variables for further training.
    \item Since the previous authors achieved the best results using Random forest, attempt to manually tune such a model with regard to $m\textsubscript{try}$.
    \item Evaluate other classification methods that may be suitable and pick the best-performing for further optimizations.
\end{itemize}

\subsection{Incorporation of Parent Commits} \label{ssec:parent_commits}

Research Question 4 was conceived in a way that allows us to validate the results as obtained in the previous questions. There, the results compare cross- vs. single-project commit classification, the accuracy of density- and size features, and how well the attributes of our extended dataset perform, also in comparison with the datasets of \citet{levin2017commits}.

To build a dataset that is made up of chains of commits, where the nature of the youngest child commit (the principal commit) is to be predicted, one needs to know about the direct predecessors (one or more generations of direct parents) of it. The labeled dataset from \citeauthor{levin2017commits} does not feature consecutively labeled commits. However, we have gathered such relational information within our extended dataset, which also covers all the commits from \citeauthor{levin2017commits}. That implies that all of the parent commits are sourced from our extended dataset, and therefore can only include its features (i.e., we do not have keyword- or code-change-features at our disposal). However, the principal commit is allowed to have any of the features. None of the commits involved is a merge-commit. We were stringent about excluding such, as those need to be further investigated due to their potentially mixed nature.

\begin{table*}
    \centering
    \begin{tabular}{|l|l|}\hline
        \thead[l]{Dataset} & \thead[l]{Type of the principal commit} \\ \hline
        A & Using the features of \citet{levin2017commits} (keywords, code-changes). \\ \hline
        B & Using only density- and size related features from our extended dataset~\cite{honel2019commits}. \\ \hline
        C & Using both the features of datasets A and B (keywords, code-changes, density-/size features). \\ \hline
        D & Same as C, but without keywords. \\ \hline
    \end{tabular}
    \caption{Types of principal commits used in the four datasets of RQ 4.}
    \label{tab:princ_commits}
\end{table*}

We are building four different datasets, which are distinguished from each other by the type of principal commit. We are differentiating four types of principal commits (cf.\ Table~\ref{tab:princ_commits}). Then for each dataset, a sub-dataset is built, featuring one or more generations of parents. We are selecting $\{1,2,3,5,8\}$ as the amounts of parents to be included, as those still yield a respectable dataset size (almost 900 commits have eight parents) and resemble the Fibonacci series. In total, we are thus using 20 datasets for Research Question 4. Since the relation of each commit to its project is retained, we can use the same datasets for cross- and single-project classification.


Research Question 4 is addressed by comparing the accuracy across its A, B, C, and D datasets. The second part is partially covered as well. However, we intend to point out the importance of the net- and gross-attributes separately. Lastly, the accuracy and Kappa of the champion models built across projects and for every single project are evaluated. For statistical models, we are using RFE based on Random forests and repeated cross-validation, using many folds to find champion models.

\section{Results}\label{sec:result}
The results are laid out for each research question and summarized at the beginning of each subsection.

\subsection*{RQ1: The Statistical Properties of the Extended Dataset}

\begin{mdframed}[nobreak=true]
    $\sbullet$~The evolutionary patterns using file- or line-counts are significantly different, contrary to prior research. \\
    $\sbullet$~There is a significant shift in what constitutes a small commit; every tenth commit affects 3--4 lines, every other already 25--50. \\
    $\sbullet$~Instead of \emph{corrective}, many zero- or near-zero size commits need to be considered as \emph{perfective} instead. \\
    $\sbullet$~Most commits have a high density. This affects commits of all sizes.
\end{mdframed}

\citet{herraiz2006comparison} found that the evolutionary patterns for commits in open source software are the same, regardless of whether they were based on counting the number of files or lines of code. \citet{purushothaman2005toward} investigate what they consider to be tiny commits, and found that 10\% of all commits are small changes, i.e., only affecting a single line.

\begin{figure}[t]
    \centering
    \input{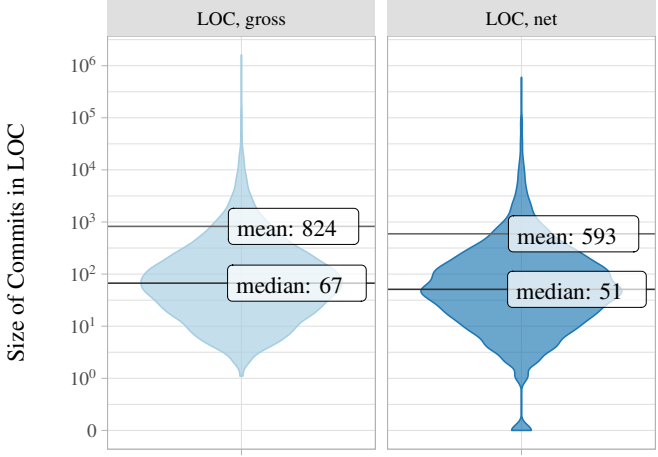}
    \vspace{-10pt}
    \caption{The size of commits in LOC, across the almost 360\,000 commits of the extended dataset.}
    \label{fig:density_net-vs-gross}
\end{figure}

To address Research Question 1, we gathered descriptive statistics for the datasets used. We find a weak correlation between notions of size based on either amount of affected files or lines of code. This suggests that it is worthwhile to investigate a commit's nature using a LOC-based notion of size. We then confirm previous results, relating the size to the nature, and find that corrective commits are usually the smallest. Research Question 1B investigates the nature of commits that change only a few lines. While we observe a difference between net- and gross-size, we find that just a small ratio of commits affects five or fewer lines. When examining the maintenance activities of such small commits further, we find that more commits should be considered perfective, using a net-notion of size.

It is important to understand the significance of the size of a commit, and especially its density. To demonstrate it, we took the size attributes of the extended dataset of commits, holding almost $360\,000$ commits, into account. Out of these, $3\,279$ ($1$\%) had a size of zero. Those are due to, e.g., starting or stopping to track files that are empty or changes to binary files that result in no lines changed. Another $98\,869$ commits had a density of one, meaning that all affected lines contributed to its net-size. $71.59$\% of the commits hence were non-empty, and had a density in the range $[0,1)$, as can be seen in Figure~\ref{fig:density_net-vs-gross}. Following our expectation, the correlation between net- and gross-counts of LOC in these commits is $0.9885$ (strong positive).

\begin{figure}[t]
    \centering
    \input{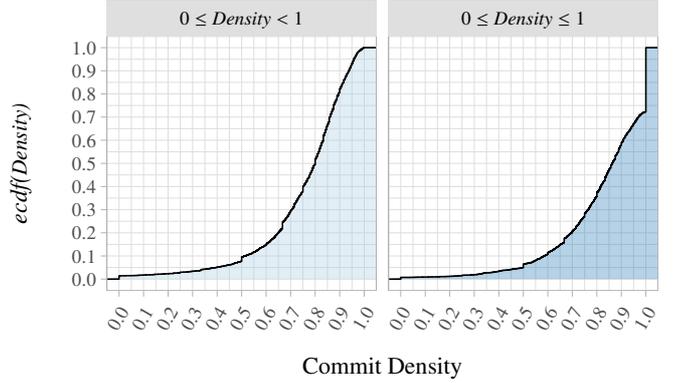}
    \caption{The empirical cumulative probability distribution of commit density across the almost 360\,000 commits of the extended dataset. About every tenth commit has a density of $0.5$ or less; about every other commit has a density of $0.8$ or less. Considering $\textit{Density}=1$ commits (right), more than every fourth has a density of $1$.}
    \label{fig:density_ecdf}
\end{figure}

\begin{figure}[t]
    \centering
    \input{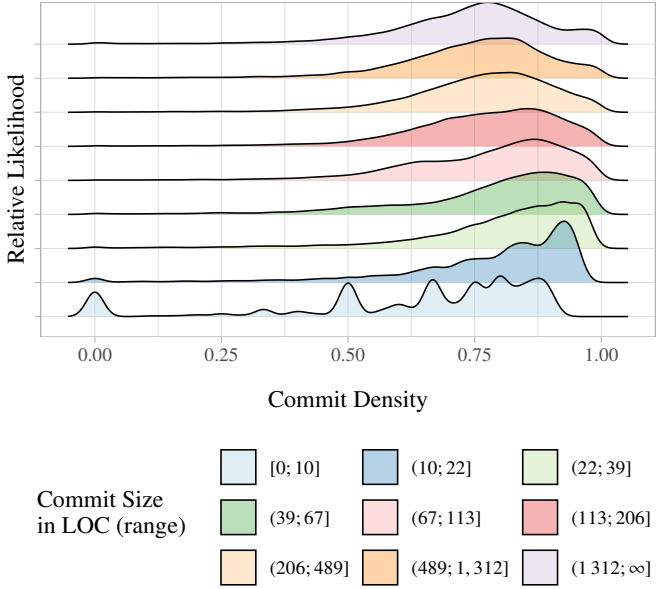}
    \vspace{-30pt}
    \caption{Ridge-plot of the probability distribution of commits' densities, for a number of delimited ranges, as expressed by gross-size lines of code. The ranges were delimited using the following quantiles: $12.5$\% ($10$), $25$\% ($22$), $37.5$\% ($39$), $50$\% ($67$), $62.5$\% ($113$), $75$\% ($206$), $87.5$\% ($489$), $95$\% ($1\,312$).}
    \label{fig:density_in_quantiles}
\end{figure}

Taking the distribution of the commit density into account, it is apparent that larger densities are much more common than lower densities. About only every 10\textsuperscript{th} commit has a density of $0.5$ or less, while already about every other commit has a density of $0.8$ or less, cf.\ Figure~\ref{fig:density_ecdf}.
To further understand commit density and how it relates to a commit's gross size, we prepare a few ranges and visualize the distribution of densities, cf.\ Figure~\ref{fig:density_in_quantiles}.

\begin{figure}[t]
    \centering
    \input{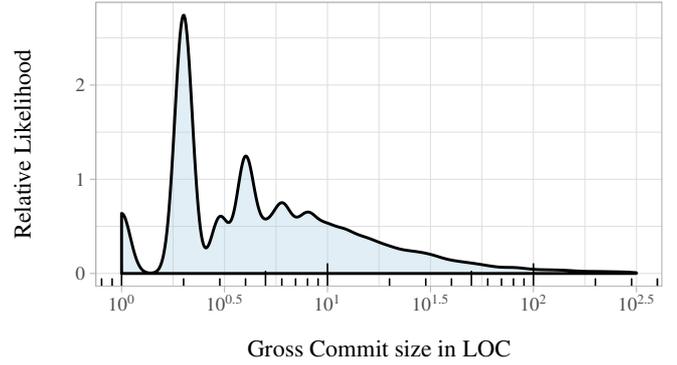}
    \vspace{-10pt}
    \caption{The distribution of commits having a density of $1$, w.r.t. their gross-size in LOC (tail truncated).}
    \label{fig:density_1_density}
\end{figure}

A significant portion of the examined commits in the extended dataset, more than $28$\%, have the maximum density of $1$, meaning that the net-size is the same as the gross-size. Hence, all lines in these commits are considered useful. Therefore, we have considered these separately. This phenomenon affects commits of all sizes, starting from one line up to several hundreds of thousands of lines. However, the majority of such commits has about ten lines or less, cf.\ Figure~\ref{fig:density_1_density}.

\begin{figure}[t]
	\centering
	\input{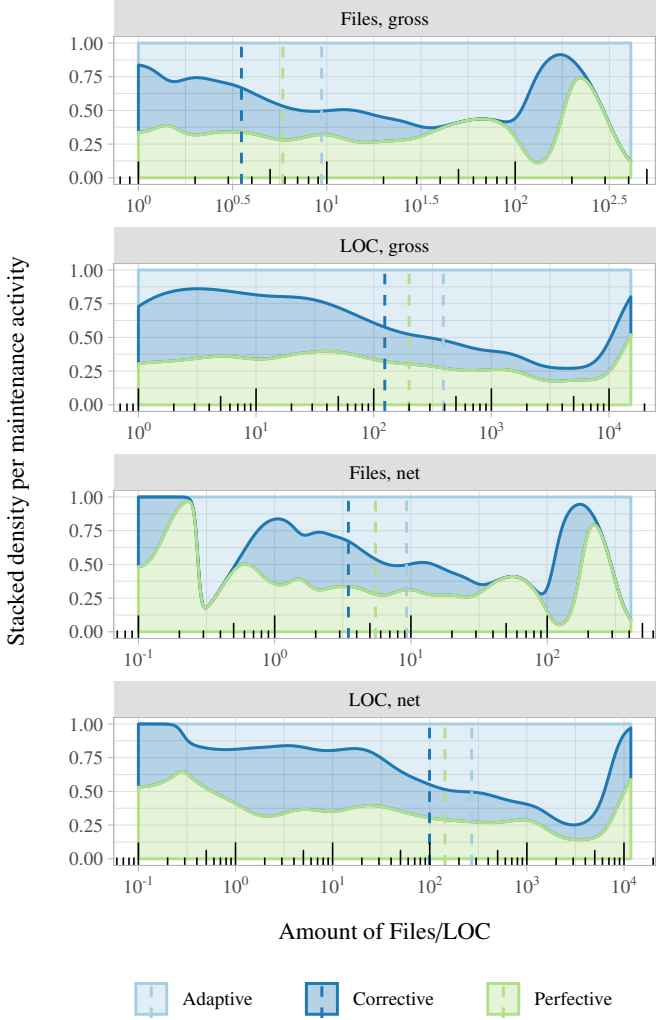}
	\vspace{-10pt}
    \caption{Density plots of net-/gross amount of files and LOC, across all projects and labels, including the mean for each label.}
	\label{fig:density_plots}
\end{figure}

\MakeUppercase{\emph{\textbf{Part A}}} Research Question 1A seeks to validate whether the size of a commit in terms of affected files or LOC is different. Prior research~\cite{herraiz2006comparison} found the difference to be insignificant. When comparing the density plots in Figure~\ref{fig:density_plots}, we observe that the minimum values for gross values (i.e., LOC or files) are $1.0$ (as a commit cannot comprise an empty set of changes), whereas the net-values can be, and in fact are, $0$. We have shifted all net-values for these plots by $0.1$, so we can use a logarithmic scale. This allows us to observe commits assigned to any of the maintenance activities, that have in fact a size of zero. Refer to Table~\ref{tab:density_plots_numeric} for the numerical properties of the various notions of size and empirical probabilities. The table also outlines the probabilities of finding commits with a size of $0$ for each of the maintenance labels.

The evolutionary patterns between classification using files and LOC are quite different. While the amount of \emph{adaptive} commits increases with size for LOC-based notions, file-based notions attribute most large commits towards \emph{perfective} and \emph{corrective} commits. The latter has its largest commits attributed to \emph{adaptive} activities, while the former identifies the largest commits to be \emph{perfective} and \emph{corrective}. Evolutionary patterns between net- and gross-size notions differ only slightly. However, we can get insights into the densities of net-notions that feature empty commits. Those could explain the differences in density when compared to their gross-sized counterparts.

\begin{table*}
    \centering
    \begin{tabular}{|l|r|r|r|r|}
        \hline
                &   \multicolumn{2}{c|}{\thead{Files}}   &   \multicolumn{2}{c|}{\thead{LOC}} \\ \hline
                &   \thead[r]{Gross}   &   \thead[r]{Net} &   \thead[r]{Gross}   &   \thead[r]{Net}     \\ \hline
        $mean\textsubscript{adaptive}$      &   $9.386$        &   $9.288$    &   $390.297$     &   $270.255$   \\
        $mean\textsubscript{corrective}$    &   $3.524$        &   $3.480$    &   $124.172$     &   $99.458$    \\
        $mean\textsubscript{perfective}$    &   $5.843$        &   $5.496$    &   $199.623$     &   $143.439$   \\ 
        $mean\textsubscript{\{a,c,p\}}$     &   $5.592$        &   $5.431$    &   $207.612$     &   $151.452$   \\ \hline
        $median\textsubscript{\{a,c,p\}}$   &   $2.000$        &   $2.000$    &   $45.000$      &   $33.000$     \\ 
        $min\textsubscript{\{a,c,p\}}$      &   $1.000$        &   $0.000$    &   $1.000$       &   $0.000$      \\ 
        $max\textsubscript{\{a,c,p\}}$      &   $411.000$      &   $411.000$  &   $15318.00$    &   $11766.060$  \\ \hline
        $P(a\, | \, x < 1)$                 &   \emph{n/a}     &   $0.000$    &   \emph{n/a}    &   $0.000$  \\      
        $P(a\, | \, 1 \leq x < 2)$          &   $0.179$        &   $0.179$	  &   $0.012$       &   $0.012$     \\
        $P(a\, | \, 2 \leq x < 5)$          &   $0.528$        &   $0.528$	  &   $0.049$       &   $0.061$     \\ \hline
        $P(c\, | \, x < 1)$                 &   \emph{n/a}     &   $0.004$    &   \emph{n/a}    &   $0.004$     \\
        $P(c\, | \, 1 \leq x < 2)$          &   $0.390$        &   $0.388$	  &   $0.010$       &   $0.020$     \\
        $P(c\, | \, 2 \leq x < 5)$          &   $0.818$        &   $0.816$	  &   $0.154$       &   $0.188$     \\ \hline
        $P(p\, | \, x < 1)$                 &   \emph{n/a}     &   $0.005$    &   \emph{n/a}    &   $0.005$     \\
        $P(p\, | \, 1 \leq x < 2)$          &   $0.340$        &   $0.357$	  &   $0.007$       &   $0.030$     \\
        $P(p\, | \, 2 \leq x < 5)$          &   $0.742$        &   $0.752$	  &   $0.101$       &   $0.134$     \\ \hline
    \end{tabular}
    \caption{Numerical properties and empirical probabilities of gross- and net datasets w.r.t. the commit's label.}
    \label{tab:density_plots_numeric}
\end{table*}

From the empirical probabilities in Table~\ref{tab:density_plots_numeric}, we can derive some statements. First, the probability of observing an empty commit of activity \emph{adaptive} is zero. If any, then the observed type of commit is either of corrective or adaptive nature. We can see that there are, occasionally, significant differences in observing either type of maintenance activity, given the size is zero or in the interval $[1,2)$. Therefore, when considering net-sized datasets, we can observe a shift in the distributions for the maintenance activities. This shift can also be observed when examining the density plots in Figure~\ref{fig:density_plots}. Additionally, regarding the weak correlations between files- and LOC-based gross- and net datasets ($0.347$ and $0.296$, respectively), it might be worth to investigate the nature of a commit w.r.t. affected LOC, instead of files, as done in prior studies~\cite{herraiz2006comparison, hattori2008nature, hindle2008large}.

\begin{figure}[t]
    \centering
    \input{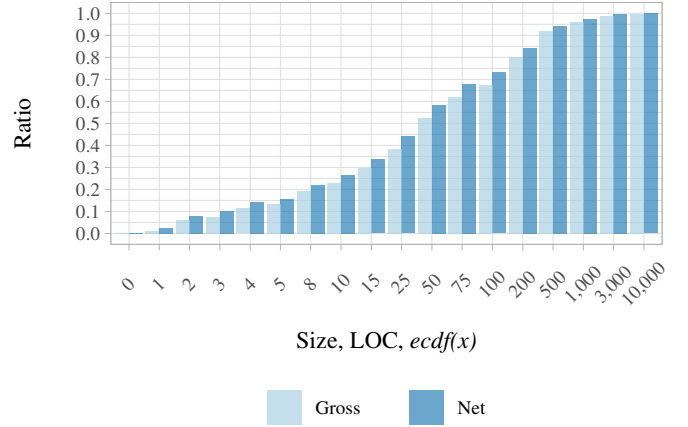}
    \vspace{-10pt}
	\caption{Empirical distribution of commits w.r.t the affected lines, separated by gross- and net size.}
    \label{fig:ratio_range_commits}
\end{figure}

Further expanding on \emph{Do development activities appear mainly in small commits?} (posed by \citeauthor{hattori2008nature}), we conclude that corrective commits are the smallest, followed by perfective, and then adaptive commits. In that same order, it is more to less likely to encounter a commit affecting between two and five lines. Our observations are, therefore, in consensus with those of \citeauthor{hattori2008nature}. The difference between gross and net is insignificant.

\MakeUppercase{\emph{\textbf{Part B}}} Research Question 1B concerns the size of commits with regard to their gross- and net-size. \citeauthor{purushothaman2005toward} found that every tenth commit changed only a single line of code and that nearly $50$\% changed ten lines or less. Given our extended dataset that spans eleven projects, we found that one out of ten commits affected about four lines or less, as can be seen in Figure~\ref{fig:ratio_range_commits}. In general, we can observe a shift towards an increased ratio of net-sized commits. More than half of the commits affected 50 lines or less in our data.

\begin{figure}[t]
    \centering
    \input{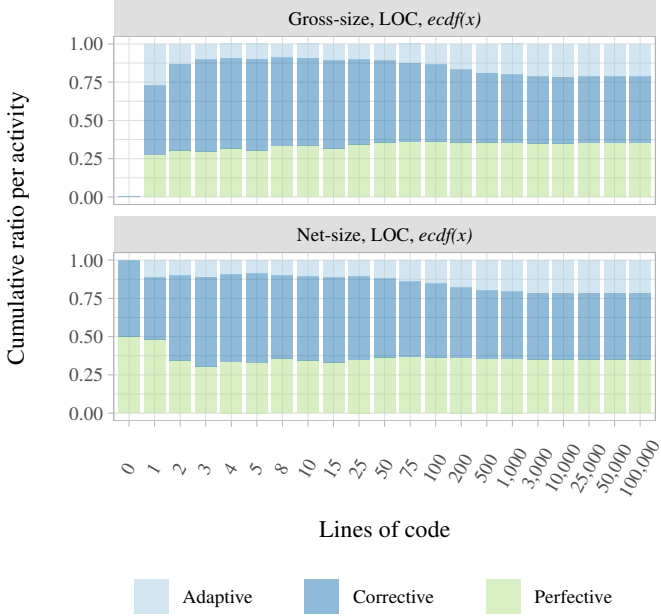}
    \vspace{-10pt}
	\caption{Empirical distribution of maintenance activities in commits, separated by gross- and net size.}
    \label{fig:ecdf_ratio_per_label}
\end{figure}

The distribution of maintenance activities across commits up to a specific size is different for gross- and net-size, as depicted by Figure~\ref{fig:ecdf_ratio_per_label}. In the upper plot, more than $25$\% of the commits are considered to be adaptive when only one line is affected by them. Regardless of the examined sizes, about $25$\% of the commits are of perfective nature. The lower plot, which depicts high-density commits, finds that commits that affected zero lines in actuality are either corrective or perfective. Almost all of the zero-lines commits that were adaptive previously, need now considered to be corrective.

\pagebreak
\subsection*{RQ2: Commit Classification using only Size and Density}

\begin{mdframed}[nobreak=true]
    $\sbullet$~Classification accuracy and Kappa show a wider spread for individual projects. Less variables are required for single projects. \\
    $\sbullet$~Net-versions of attributes are deemed more important than their respective gross-counterpart. \\
    $\sbullet$~While models using net-size variables profit from each added variable, this leads to larger models. Gross-size based models are less complex and perform slightly better.
\end{mdframed}

\begin{figure*}
	\centering
	\input{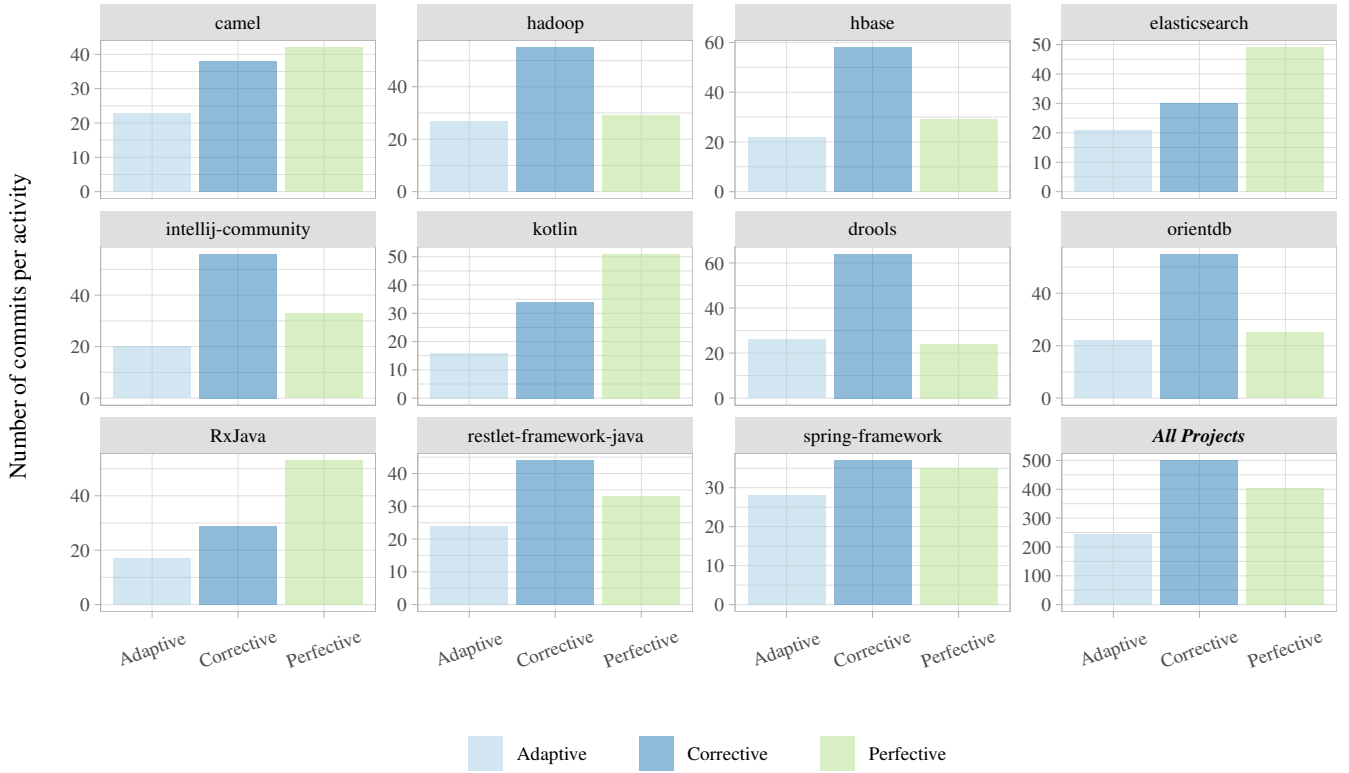}
    \caption{Distribution of commit maintenance activities across each individual project, and for all projects combined (last plot).}
	\label{fig:sp-and-xp_labelDist}
\end{figure*}

We have found differences in accuracy between classifiers trained across the entire dataset and for each project individually. This is also partly due to the different distributions of maintenance activities in each project, esp. when compared across all projects. In Figure~\ref{fig:sp-and-xp_labelDist} the most common maintenance activity across projects is \emph{corrective}. However, for individual projects, we observe that this is not always the most prevailing activity. By correlating the attributes of our extended dataset, we find strong positive correlations and can prune it considerably. It becomes clear that net-versions of attributes are deemed more important than their respective gross-counterpart.

\begin{figure}[t]
	\centering
	\input{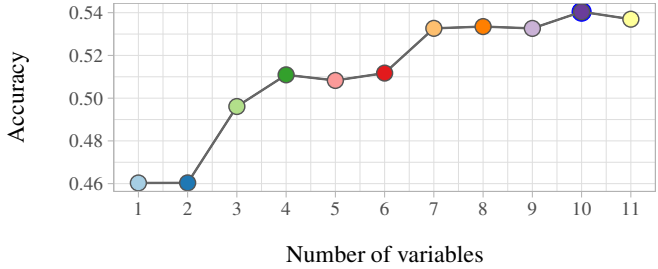}
	\vspace{-10pt}
    \caption{Cross-project model training performance w.r.t. number of most important variables. Each point represents a champion model based on the amount of variables denoted. The resampling during training was done using a five times repeated, 10-fold cross-validation.}
	\label{fig:model_perf_numVars}
\end{figure}

\begin{figure}[t]
    \centering
    \input{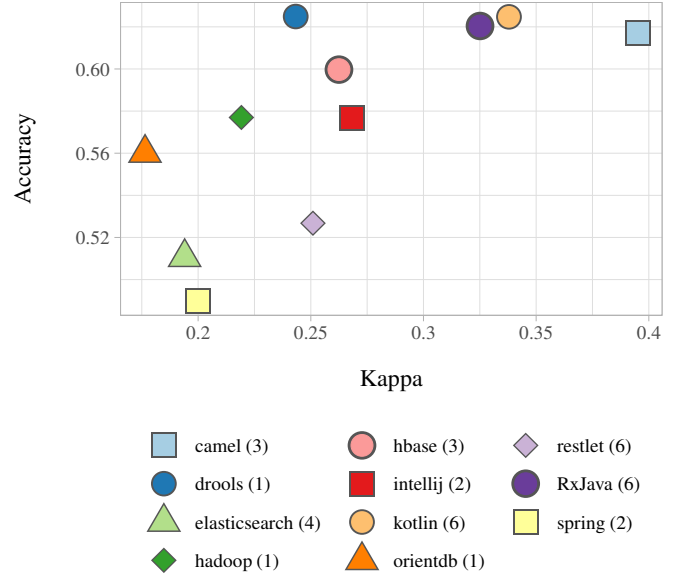}
    \vspace{-10pt}
	\caption{Accuracy and Kappa for each project. The number in parentheses represents the number of variables used.}
    \label{fig:acc_Kappa_per_project}
\end{figure}

\MakeUppercase{\emph{\textbf{Part A}}} We use the methods described in section \ref{ssec_stat_meth_rq2a} to perform a series of experiments. From this, we report the following results for Research Question 2A. Generally, classification accuracy and Kappa had a wider spread for individual projects (cf.\ Table~\ref{tab:acc_Kappa_vars} and Figures~\ref{fig:model_perf_numVars}~\&~\ref{fig:acc_Kappa_per_project}), compared to cross-project classification.

Our tool, \emph{\mbox{Git-Density}}~\cite{honel2019gitdensity}, adds 22 size attributes (cf.\ subsection~\ref{ssec_dataset}). Using a correlation coefficient of $0.75$, we have identified eleven highly correlated attributes (Number of Files Added Gross, Number of Files Deleted Net, Number of Files Renamed Gross, Number of Files Modified Gross, Number of Lines Added by Added Files Gross, Number of Lines Deleted by Deleted Files Net, Number of Lines Added by Modified Files Gross, Number of Lines Deleted by Modified Files Gross, Number of Lines Added by Renamed Files Gross, Number of Lines Added by Renamed Files Net, Number of Lines Deleted by Renamed Files Gross) across projects.

Attribute correlation across projects kept eleven attributes, nine of which were the \emph{net}-version of an attribute; the attributes \emph{Density} and \emph{Affected Files Ratio Net} were kept. For the eleven single projects, we instead counted the most highly correlated attributes for each project. Those were (followed by the count) Number of Lines Deleted by Deleted Files Gross (5), Number of Files Deleted Gross (4), Number of Files Renamed Net (4), Number of Lines Added by Added Files Net (4), Number of Lines Deleted by Renamed Files Net (4), Number of Files Added Net (2), Number of Lines Added by Modified Files Net (2), and Number of Lines Deleted by Modified Files Net (2).

Since correlation is always done pairwise, the variable with the largest mean absolute correlation is identified for removal. Note that we have not removed the highly correlated attributes, though, as the next step was applying the variable importance, which determines the importance of each predictor independent of the correlation. 

We can report low variance for features derived from deleted or renamed files. This is somewhat expected, as deleting and renaming files occurs much more infrequent than adding new or modifying existing files. The overall \emph{Density} attribute we engineered shows high variance and is the fifth most important predictor (out of eleven) across all size attributes, with average and maximum importances of $61.91$\% resp. $72.51$\% across projects.

\begin{figure}[t]
    \centering
    \input{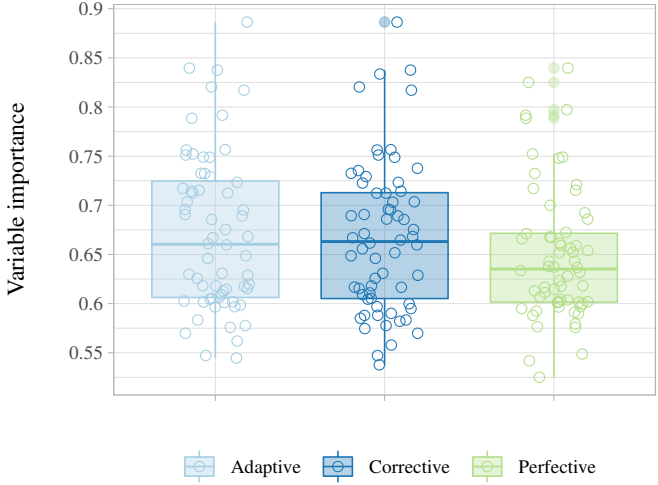}
	\vspace{-10pt}
	\caption{Boxplots for variable importances for single projects, per maintenance activity. Slight differences for predicting each label from our extended dataset can be observed.}
    \label{fig:acp_boxplots}
\end{figure}

Since there is less data available to examine the variable importance for each single project, it frequently happened that specific attributes showed no variance within the scope of a project. The seven attributes that consistently remained across all projects were Number of Lines Added by Modified Files Net, Number of Lines Added by Added Files Net, Number of Files Added Net, Number of Files Modified Net, Density, Number of Lines Deleted by Modified Files Net, and Affected Files Ratio Net (ordered descending by average importance across labels). Note that all of these attributes are the net-version of their feature.
We have further examined the importance of each remaining attribute for each of the maintenance labels \emph{\{a,c,p\}} separately. It is noteworthy that our engineered feature \emph{Affected Files Ratio Net} improved considerably by $45$\% (averaging at $58.37$\%) and that \emph{Density} gained about $3$\% in importance. Refer to Figure~\ref{fig:acp_boxplots} for detailed boxplots.

\begin{table}
    \centering
    \begin{tabular}{|l|r|r|}
        \hline
        \thead[l]{Measurement} & \thead[r]{Cross-project} & \thead[r]{Individual Project} \\ \hline
        Accuracy, ZeroR & $0.435$   & $0.370$--$0.561$ \\ \hline
        Accuracy, best & $0.547$ & $0.652$   \\
        Accuracy, worst & $0.450$ & $0.462$   \\ \hline
        Kappa, best & $0.267$   & $0.395$   \\
        Kappa, worst & $0.092$  & $0.051$  \\ \hline
        Variables, best     &   10   &   6   \\
        Variables, worst    &   2   &   5   \\ \hline
    \end{tabular}
    \caption{Best/worst Accuracy, Kappa and number of variables cross- and per project.}
    \label{tab:acc_Kappa_vars}
\end{table}

Model-selection for single projects shows peculiarities for the projects \emph{Drools} and \emph{Hadoop}, where the best model uses only one variable (Number of Files Modified resp. Number of Lines Added by Added Files Net). This is somewhat surprising but likely explained by underfitting, as there is only a low amount of commits available per project---the other projects used between two and six variables each. The best and worst accuracy for single projects, however, is greater than for cross-project classification. The range for Kappa is considerably larger and reaches higher absolute values (values between $0.21$ and $0.4$ are considered \emph{fair} \cite{landis1977application}). Also, the amount of variables required is noticeably lower for classification in single projects, cf.\ Table~\ref{tab:acc_Kappa_vars}. We applied the ZeroR classifier to obtain a baseline for classification performance. Across projects, all trained models performed better than it. For individual projects, the best-performing models achieved an accuracy that was higher by $7.54$\% on average, compared to ZeroR. Kappa is used to  measure the chance-corrected agreement between the model's predicted classifications and the true labels. It is an important measure because the number of available labels per class differ (cf.\ Figure~\ref{fig:sp-and-xp_labelDist}).

\begin{table}
	\centering
	\begin{tabular}{|l|c|r|r|r|r|}
		\hline
		\thead[l]{Model Type} & \thead{\# of Vars.} & \thead[r]{Acc.} & \thead[r]{Kappa} & \thead[r]{Acc., SD} & \thead[r]{Kappa, SD} \\ \hline
		gross	&   $9$		& $0.556$	& $0.280$		& $0.031$	&	$0.057$	\\ \hline
		net		&	  $10$	& $0.547$	& $0.265$		& $0.038$	&	$0.061$	\\ \hline
	\end{tabular}
	\caption{Comparison of the training performance of the best cross-project models for net- and gross based datasets.}
	\label{tab:net_vs_gross_models}
\end{table}

\MakeUppercase{\emph{\textbf{Part B}}} As for resolving Research Question 2B, we have also examined the classification accuracy and Kappa, both cross- and per-project, then net- vs. gross-size (cf.\ Table~\ref{tab:acc_Kappa_net_gross}). The most significant result is an accuracy of $65$\% and Kappa of $0.39$ for a single project, using only size data. Again, we obtain better results when training models on a per-project basis, rather than attempting cross-project classification. As for the remaining results, we follow the laid out methods of subsection~\ref{ssec_stat_meth_rq2b}.

Both datasets feature ten size attributes, and we have also retained the features Affected Files Ratio Net and Density for the net-dataset, yielding twelve attributes for the latter. Because of this comparatively low amount of variables, all possible model sizes were tested using an RFE-approach. The gross-based model performs best using six variables (with no further improvement using up to ten variables), while the net-based model continually improves with each added variable, thus also using all twelve available predictors. The net-based models perform insignificantly worse, see Table~\ref{tab:net_vs_gross_models} for a complete comparison of the best models per type. The baseline for each model to outperform is set at $43.52$\% using ZeroR.

\begin{table}
    \centering
    \begin{tabular}{|l|r|r|r|r|}
        \hline
        \thead[l]{Aggregation} & \thead{Acc., net} & \thead{Acc., gross} & \thead{Kappa, net} & \thead{Kappa, gross} \\ \hline
        max &	$0.547$	& $0.556$	& $0.265$		& $0.280$ \\
        min	&   $0.455$	& $0.439$	& $0.095$		& $0.065$  \\
	    avg	&   $0.514$	& $0.519$  & $0.208$		& $0.215$  \\  \hline
        \multicolumn{5}{c}{$\uparrow$ cross-project, single-project $\downarrow$} \\
        \hline
        max	&   $0.652$  & $0.640$	& $0.395$	& $0.331$ \\
        min	&   $0.462$  & $0.472$	& $0.051$	& $0.034$ \\
        avg	&   $0.565$  & $0.566$	& $0.234$	& $0.232$ \\ \hline
    \end{tabular}
    \caption{Best/worst Accuracy \& Kappa, net vs. gross, cross- and single-projects.}
    \label{tab:acc_Kappa_net_gross}
\end{table}

Given the results from Table~\ref{tab:net_vs_gross_models}, the gross-based models should be preferred, as those have a slightly lower complexity, due to the lower amount of variables, while achieving marginally better results as their net-based counterparts. With roughly $12$\% better performance as compared to the classification results of ZeroR, these models demonstrate their significance.

\subsection*{RQ3: Size and Density for advancing the State of the Art}

\begin{mdframed}[nobreak=true]
    $\sbullet$~We can successfully reproduce previous results using Random forests. \\
    $\sbullet$~Extending previous models without selecting most important size attributes leads only to a marginal improvement. \\
    $\sbullet$~Observations suggest the interchangeability of $model\textsubscript{changes}$ and $model\textsubscript{density}$.
\end{mdframed}

We were successful in reproducing the previous authors' results, achieving outcomes very similar. Involving size attributes when comparing model performances, a slight improvement of about $2-3$\% in accuracy can be observed during training. Also, it seems that the relatively expensive to obtain change-features can be replaced by the density-features, without a decline in accuracy. By attempting further tuning and pruning and facilitating additional classifiers, we were then able to boost accuracy to up to $89$\% with a Kappa of $0.82$.

\MakeUppercase{\emph{\textbf{Part A}}} The third research question is three-part. Overall, we are interested in whether the additional size features can advance state of the art in commit classification. We could choose to establish our own baseline or, better, to reproduce the results from \citet{levin2017boosting}. We chose the latter, as this validates their results and provides comparability. They use a compound model that is based on one or two sub-models (left and right), which are of either kind $model\textsubscript{keywords}$, $model\textsubscript{changes}$, or $model\textsubscript{combined}$, where the latter is trained using all of their originally available features. If the compound model only use a single type of sub-model, that same model is used on the left and the right side. The left and right sides have a meaning for the classifier, so that the two configurations $\{model\textsubscript{A},model\textsubscript{B}\}$ and $\{model\textsubscript{B},model\textsubscript{A}\}$ are treated distinctly. The custom routine for classification using compound models chooses the left model whenever a commit uses any of the keywords the $model\textsubscript{keywords}$ was trained with. For details about the models, refer to \citet{levin2017boosting}.

While the previous authors have examined classification performance using J48-, GBM- and Random forest-classifiers, we chose to only reproduce these results using the latter for Research Question 3A, as their best performing model used that. We obtain the original dataset and perform an 85/15 percentage split, thereby withholding the smaller partition entirely from training. We then vertically split the dataset into $ds\textsubscript{keywords}$ and $ds\textsubscript{changes}$ (the entire width of the dataset is needed for the $model\textsubscript{combined}$). Then, using five times repeated 10-fold cross-validation, we train the three aforementioned models separately.

First, we are interested in training performance. To assess it, we combine the numeric votes for each class by either model and select the highest (i.e., the most probable predicted label). The training results are reported in Table~\ref{tab:org_train_performance}. We get similar results with regard to training performance, except for models \#7 and \#8, which perform significantly better (about $+15$\% improved accuracy and additional Kappa of $0.25$). As for the performance using the validation samples, our results are again similar (cf.\ Table~\ref{tab:org_valid_performance}). Surprisingly, we obtain the best results with a keywords-only compound model.

According to these results, our champion compound model is \#1; it performs slightly better than  model \#5, which was the best model for \citeauthor{levin2017boosting}. When passing the $15$\% of previously unseen validation samples through those trained models, we get similar results. Model \#5 is only insignificantly worse than model \#4, and performs almost as well as that from \citeauthor{levin2017boosting} ($76.7$\% with Kappa of $0.635$). Again, our results may be within the margin of error.

\begin{table}
	\centering
	\begin{tabular}{|l|l|l|r|r|}
	    \hline
	    \thead[l]{Model \#} &  \thead[l]{$model\textsubscript{left}$} & \thead[l]{$model\textsubscript{right}$} & \thead[r]{Accuracy} & \thead[r]{Kappa} \\ \hline
	    1   &   Combined    &   Combined    &   $0.730$    &   $0.579$ \\ \hline
	    2   &   Combined    &   Keywords    &   $0.728$    &   $0.576$ \\ \hline
	    3   &   Combined    &   Changes     &   $0.685$    &   $0.501$ \\ \hline
	    4   &   Keywords    &   Changes     &   $0.673$    &   $0.478$ \\ \hline
	    5   &   Keywords    &	Combined    &	$0.728$    &   $0.576$ \\ \hline
	    6   &	Keywords    &	Keywords    &	$0.716$    &	$0.559$ \\ \hline
	    7   &	Changes     &	Combined    &	$0.685$    &   $0.501$ \\ \hline
	    8   &	Changes     &	Keywords    &	$0.673$    &	$0.478$ \\ \hline
	    9   &	Changes     &	Changes     &	$0.527$    &	$0.248$ \\ \hline
	\end{tabular}
	\caption{Training performance of all nine compound models using the original dataset.}
	\label{tab:org_train_performance}
\end{table}

\MakeUppercase{\emph{\textbf{Part B}}} For the second part of this question, we are adding one model trained on size data. Also, $model\textsubscript{combined}$ is extended with those features. The list of compound models is extended by seven additional models in the following way:

\begin{itemize}
    \item Add a compound model for each of the other model types $\{\textit{keywords}, \textit{changes}, \textit{combined}\}$ with the $model\textsubscript{density}$ as the left model.
    \item Create three additional compound models, with $model\textsubscript{density}$ as the right model.
    \item Add a purely density compound model, where the left and the right models are both of type $model\textsubscript{density}$.
\end{itemize}

As for the baseline, the models need to beat $43.45$\% accuracy during training and $43.86$\% during validation, to be significant. As expected, the density-only compound model performs exactly as in the previous research question. The best such compound model consists of $model\textsubscript{density}$ and $model\textsubscript{combined}$, achieving $64.62$\% accuracy with a Kappa of $0.427$ during training. The performance of the other compound models that include the $model\textsubscript{combined}$ (which now spans size features) declines on average by $2$--$3$\% accuracy during training.

\begin{table}
	\centering
	\begin{tabular}{|l|l|l|r|r|}
	    \hline
	    \thead[l]{Model \#} &  \thead[l]{$model\textsubscript{left}$} & \thead[l]{$model\textsubscript{right}$} & \thead[r]{Accuracy} & \thead[r]{Kappa} \\ \hline
	    4   &   Keywords    &   Changes     &   $0.731$    &   $0.573$ \\ \hline
	    5   &   Keywords    &	Combined    &	$0.766$    &   $0.631$ \\ \hline
	    6   &   Keywords    &	Keywords    &	$0.784$    &   $0.660$ \\ \hline
	\end{tabular}
	\caption{Validation performance of some selected compound models using the original dataset.}
	\label{tab:org_valid_performance}
\end{table}

Using the validation samples, the best-performing compound model now is $model\textsubscript{keywords}$, $model\textsubscript{combined}$ with an accuracy of $75.44$\% and Kappa of $0.619$, which denotes a slight improvement over the previous authors' results.

\begin{table}
	\centering
	\begin{tabular}{|l|l|l|r|r|}
	    \hline
	    \thead[l]{Model \#} &  \thead[l]{$model\textsubscript{left}$} & \thead[l]{$model\textsubscript{right}$} & \thead[r]{Accuracy} & \thead[r]{Kappa} \\ \hline
	    9   &	Changes &	Combined    &	$0.626$    &	$0.417$ \\ \hline
	    10	&   Changes &	Keywords    &	$0.608$    &    $0.388$ \\ \hline
	    11  &	Changes &	Changes     &	$0.561$    &	$0.307$ \\ \hline
	    12  &	Changes &	Density     &	$0.532$    &	$0.245$ \\ \hline
	    13  &	Density &	Combined    &	$0.626$    &	$0.403$ \\ \hline
	    14  &	Density &	Keywords    &	$0.608$    &	$0.374$ \\ \hline
	    15  &	Density &	Changes	    &   $0.561$    &	$0.290$ \\ \hline
	    16  &	Density &   Density	    &   $0.532$    &	$0.225$ \\ \hline
	\end{tabular}
	\caption{Validation performance of density- and change based models.}
	\label{tab:ext_valid_performance}
\end{table}

\begin{figure}[t]
	\centering
	\input{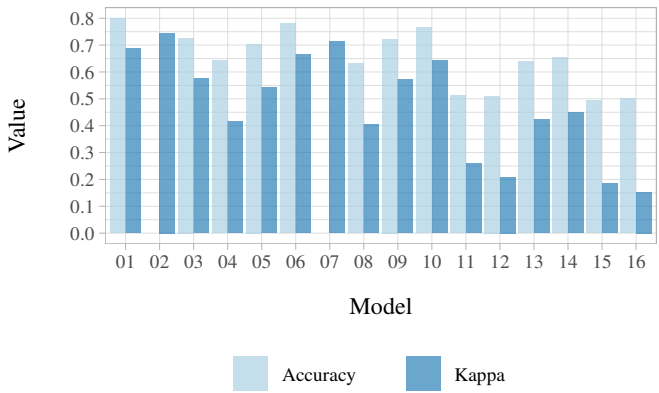}
	\vspace{-10pt}
    \caption{Performance of all 16 compound models on the validation samples.}
	\label{fig:valid_performance_ex_models}
\end{figure}

We seem to be able to swap out code-changes for density, when the respective other model is of type $model\textsubscript{keywords}$, which might be worthwhile due to the lower cost of obtaining it. When further comparing models \#9 through \#12 and \#13 through \#16, we observe similar performance with either changes- resp. density-based models on the left, which is another hint at the interchangeability of these kinds of models. Overall, we observe a drop in performance in models \#9 through \#16, all of which use either changes or density as their left model (cf.\ Table~\ref{tab:ext_valid_performance} and Figure~\ref{fig:valid_performance_ex_models}).

\MakeUppercase{\emph{\textbf{Part C}}} The last part of this question is concerned with attempting to improve the performance of a model that includes all types of original and extended attributes (keywords, changes, density). We are pruning the underlying dataset in the first place, eliminating all gross-size, zero- and near-zero-variance attributes. This step reduced the dataset to less than 35 attributes. We then ran an RFE, which yielded the best model using 26 attributes that achieves an accuracy of $70.87$\% with a Kappa of $0.544$ during training. Among the ten most important variables, we find three density attributes, four related to keywords, and three to changes.

\citeauthor{levin2017boosting} achieved significant results with Random forests, so we attempt to manually tune such a model with respect to its $m\textsubscript{try}$-parameter. However, the performance did not improve compared to the best model we found using RFE. We  tried a  mix of classifiers on the pruned dataset and trained each with five times repeated 10-fold cross-validation. The results are reported in Table~\ref{tab:tune_various_methods}. The baseline set by ZeroR is an accuracy of $43.86$\%.

\begin{table*}
	\centering
	\begin{tabular}{|l|l|r|r|}
		\hline
		\thead[l]{Method} & \thead[l]{\emph{R} package} & \thead[r]{Accuracy} & \thead[r]{Kappa} \\ \hline
		\textbf{ZeroR} (baseline)   & \emph{n/a}\footnotemark  &   $0.439$    &   $0.0$ \\ \hline
		\textbf{\citeauthor{levin2017boosting}}~\cite{levin2017boosting} (Random forest) & \emph{n/a}\footnotemark    &   $0.760$  &   $0.630$ \\ \hline

		LogitBoost (typical) & \multirow{3}{*}{\parbox[t]{2cm}{\normalsize\raggedright \emph{caTools}\footnotemark}} & $\textbf{0.805}$ & $\textbf{0.690}$ \\
		LogitBoost ($model\textsubscript{keywords}$) & & $\textbf{0.850}$ & $\textbf{0.780}$ \\
		LogitBoost ($model\textsubscript{combined}$, $model\textsubscript{keywords}$) & & $\textbf{0.891}$ & $\textbf{0.826}$ \\ \hline

		Least Squares SVM (lssvmRadial) & \multirow{3}{*}{\parbox[t]{2cm}{\normalsize\raggedright \emph{kernlab} \cite{rpack_kernlab}}} & $0.673$ & $0.482$ \\
		SVM (radial kernel) & & $0.632$    &   $0.413$ \\
		SVM (linear kernel) & & $0.713$    &   $0.554$ \\ \hline
		
		Neural Network & \multirow{2}{*}{\parbox[t]{2cm}{\normalsize\raggedright \emph{nnet} \cite{rpack_MASS}}} & $0.696$    &   $0.523$ \\
		Model Averaged Neural Network (avNNet) & & $0.708$ & $0.540$ \\ \hline

		Gradient Boosting Machine   & \emph{gbm}\footnotemark  &   $0.725$    &   $0.570$ \\ \hline
		eXtreme Gradient Boosting (xgbTree) & \emph{xgboost}\footnotemark & $0.708$ & $0.543$ \\ \hline
		Linear Discriminant Analysis (lda)  & \emph{MASS} \cite{rpack_MASS} & $0.673$ & $0.491$ \\ \hline
		Mixture Discriminant Analysis & \emph{mda}\footnotemark & $0.708$ & $0.540$ \\ \hline
		C5.0                        &   \emph{C50}\footnotemark & $0.702$ & $0.535$ \\ \hline
		Naive Bayes                 & \emph{naivebayes}\footnotemark & $0.544$ & $0.253$ \\ \hline
	\end{tabular}
	\caption{Overview of attempted methods for classification on the tuned dataset, compared to the state of the art (5 times repeated 10-fold cross-validation, results on validation samples).}
	\label{tab:tune_various_methods}
\end{table*}

\addtocounter{footnote}{-8}
\stepcounter{footnote}\footnotetext{Using an own implementation done in R to predict the most common label. This results in a Kappa of $0$.}
\stepcounter{footnote}\footnotetext{The authors did not disclose which package they were using, however, they used R as well.}
\stepcounter{footnote}\footnotetext{https://cran.r-project.org/web/packages/caTools/}
\stepcounter{footnote}\footnotetext{https://cran.r-project.org/web/packages/gbm/}
\stepcounter{footnote}\footnotetext{https://cran.r-project.org/web/packages/xgboost/}
\stepcounter{footnote}\footnotetext{https://cran.r-project.org/web/packages/mda/}
\stepcounter{footnote}\footnotetext{https://cran.r-project.org/web/packages/C50/}
\stepcounter{footnote}\footnotetext{https://cran.r-project.org/web/packages/naivebayes/}

The \emph{LogitBoost} classifier outperforms all other methods significantly, so we selected it for further tuning. The performance improves if we use the full combined-dataset instead of the one we just pruned. We have repeatedly run the training and report improved classification results. The most solid performing model uses $model\textsubscript{keywords}$ only, achieving a stable accuracy of $85$\% with a typical $0.78$ Kappa. The best results, however, were obtained using the compound model $model\textsubscript{combined}$, $model\textsubscript{keywords}$, peaking at $89.13$\% accuracy with a Kappa of $0.826$. During subsequent runs, however, that model typically dropped to $80$\% with a Kappa of $0.69$. This is likely explained by how the validation samples are selected between runs. Note that Kappa values between $0.61$ and $0.8$ are considered \emph{substantial}, and values between $0.81$ and $1.0$ are considered \emph{almost perfect}~\cite{landis1977application}.

\subsection*{RQ4: Using the Size and Density of previous generations}

\begin{mdframed}[nobreak=true]
    $\sbullet$~It is best to pick a principal commit that uses all available features. \\
    $\sbullet$~Looking back up to three commits in time improves prediction accuracy. \\
    $\sbullet$~Models trained for single projects profit significantly from considering preceding commits and achieve an accuracy beyond $93$\% with an almost perfect Kappa of $0.88$.
\end{mdframed}

For the fourth research question, we are examining three aspects in particular. First, we attempt to determine which features in a principal commit with appended parent generations are the most important. We find that using size features only performs worst. We then confirm that size features can replace the comparatively expensive code-changes features. Lastly, we remove keyword-features from the principal commit and report only an insignificant decline in accuracy.
Second, we are interested in the amount of retained variables in previous generations of the principal commit. It appears that some datasets tend to retain more features than others and that a fair amount of these retained features are size-based net-features.
Third, we examine the differences in models' performance trained across all projects and for individual projects. These results are contrasted to outcomes obtained earlier in this study, in Research Question 2. We find significant improvements for either, with new absolute champion models trained for individual projects.

\begin{figure*}[!ht]
	\centering
	\input{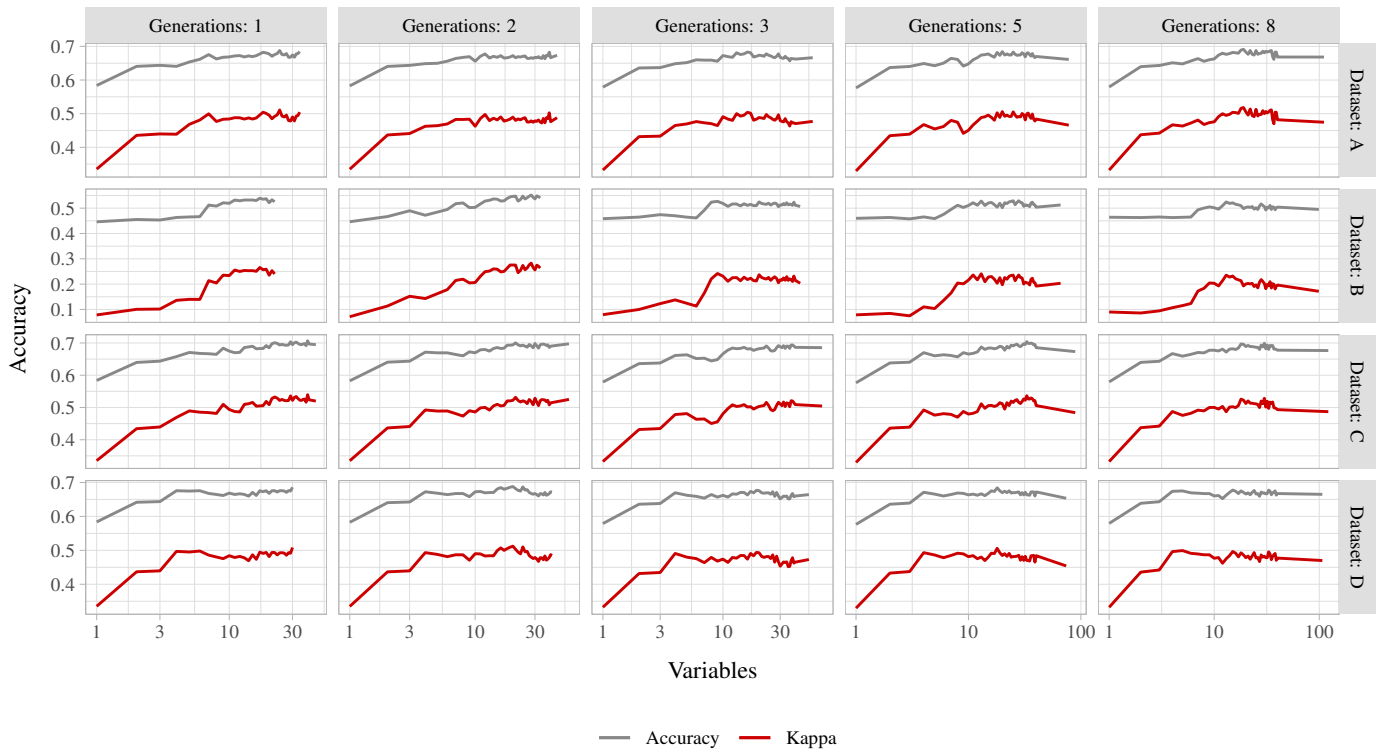}
	\vspace{-20pt}
    \caption{Cross-project model performance including up to one, two, three, five and eight (left to right) previous generations (directly preceding commits). The four rows correspond to the four datasets A, B, C and D built for research question four.}
	\label{fig:rq4_xp_AccKappa}
\end{figure*}

\MakeUppercase{\emph{\textbf{Part A}}} In Research Question 4A, we attempt to find out which type of principal commit (cf.\ Table~\ref{tab:princ_commits}) is most suitable when attaching features of commits of previous generations to evaluate prediction performance. We built the four datasets A, B, C, and D to evaluate this (cf.\ Figure~\ref{fig:rq4_xp_AccKappa}). Refer to subsection~\ref{ssec:parent_commits} for how these datasets were constructed. The worst-performing models are all based on dataset B, which features only size features.
We confirm our previous findings that replacing out code-change-features with size features does not result in a decline in model performance. This is an important finding when comparing the engineering-cost for either set of features.
Models based on the D dataset perform reliably, even though we have eliminated the keyword-features. With a slight margin over models based on the A dataset, models based on the C datasets are our champion models, for cross-project classification. Models based on the D datasets perform best for single-project classification, by a slight margin over C. C-based models contain all the features, those from~\citet{levin2017commits}, and those from our extended dataset~\cite{honel2019commits}.

\begin{table}
    \centering
    \begin{tabular}{|l|r|r|}
        \hline
        \thead[l]{Measurement} & \thead[r]{Cross-project} & \thead[r]{Individual Project} \\ \hline
        Accuracy, ZeroR & $0.432$--$0.440$   & $0.371$--$0.588$ \\ \hline
        Accuracy, best & $0.707$ (C) & $0.932$ (D)   \\
        Accuracy, worst & $0.525$ (B) & $0.300$ (B)   \\ \hline
        Kappa, best & $0.540$ (C)   & $0.882$ (D)   \\
        Kappa, worst & $0.235$ (B)  & $0.04$ (C)  \\ \hline
    \end{tabular}
    \caption{Best/worst Accuracy, Kappa and number of variables cross- and per project, involving multiple generations, for each dataset.}
    \label{tab:acc_Kappa_vars_multgen}
\end{table}

\MakeUppercase{\emph{\textbf{Part B}}} Research Question 4B is concerned with the importance of features retained from preceding generations. While we can observe a positive trend for both accuracy and Kappa, up to and including three generations, this trend seemingly becomes negative beyond that or at least stagnates. In other words, looking back more than three commits to confidently predict the label of the principal commit is not of value (cf.\ Figure~\ref{fig:rq4_xp_AccKappa}).

\begin{figure*}[!ht]
	\centering
	\input{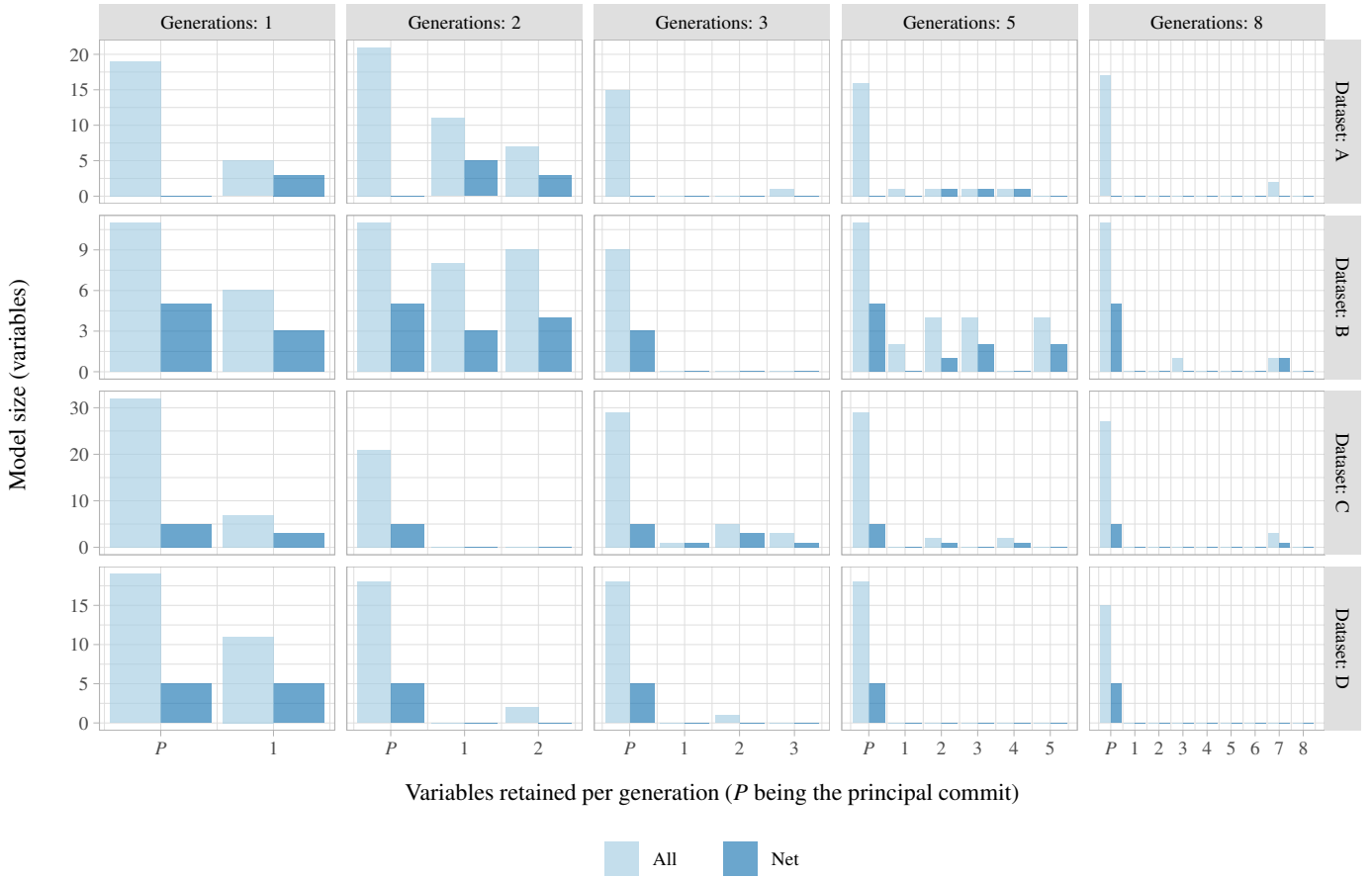}
	\vspace{-10pt}
    \caption{Amount of retained (net-)variables in each generation, including up to eight generations. The four rows correspond to the four datasets A, B, C and D built for research question four.}
	\label{fig:rq4_xp_retainedVars}
\end{figure*}

This figure is accompanied by Figure~\ref{fig:rq4_xp_retainedVars}. In it, for each sub-dataset, the size (in terms of variables) of the champion-model is shown. Each model was computed using RFE. Recall that, while the principal commit may exhibit various features, the commits from preceding generations only contain size features from our extended dataset. For the B dataset, which is based on extended features only, the most size features are retained across many-generation models. The amount of retained net-variables in previous generations is fair and is approximately one-third of the total amount of variables. Models based on datasets A and C appear to retain the second-most features across generations, whereas models based on D datasets do not appear to incorporate features of previous generations well.

\begin{figure}[t]
	\centering
	\input{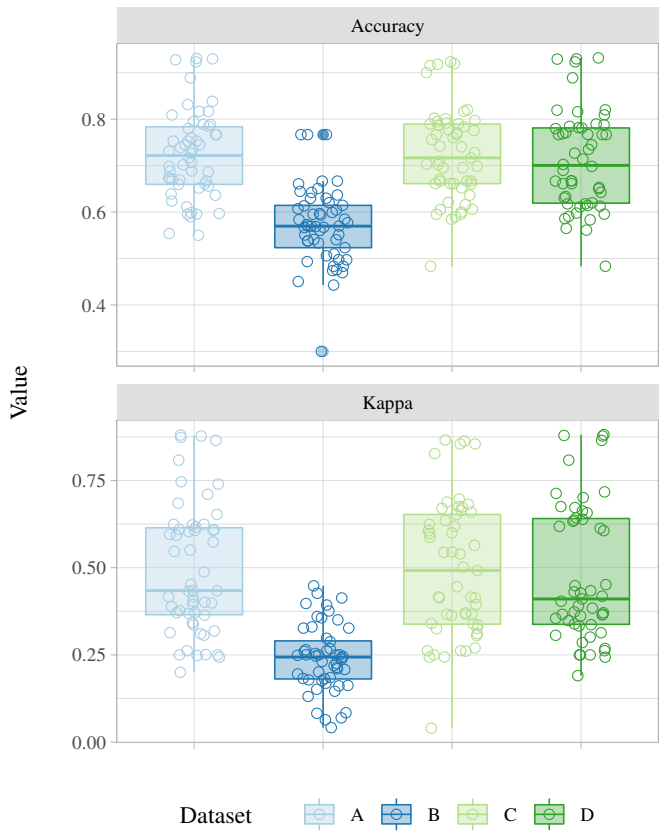}
	\vspace{-10pt}
    \caption{Aggregated results of the best models trained on individual projects for attempting an RFE in each dataset, across many generations.}
	\label{fig:rq4_sp_AccKappaBoxPlots}
\end{figure}

\MakeUppercase{\emph{\textbf{Part C}}} In the last part, we contrast these results to those of Research Question 2 (cf.\ Table~\ref{tab:acc_Kappa_vars}). Due to the vast amount of results, when introducing datasets specific to a type of principal commit and generations, we decided to aggregate them (cf.\ Table~\ref{tab:acc_Kappa_vars_multgen}). 
While the ZeroR accuracy differs only slightly when comparing to the results of Table~\ref{tab:acc_Kappa_vars} to those of Table~\ref{tab:acc_Kappa_vars_multgen}, we are reporting significant improvements for accuracy ($+15.98$\% resp. $+27.97$\%) and Kappa ($+0.273$ resp. $+0.486$), both for cross- and single-project classification for the trained champion models, respectively. We observe a decline for the worst values in single-project classification (worst accuracy $-16.19$\% and worst Kappa $-0.01$). However, models trained for single projects with an accuracy beyond $93$\% with an \emph{almost perfect} Kappa of $0.88$ allows for commit classification with great confidence. The obtained results for accuracy and Kappa for cross-project evaluations are shown in Table~\ref{tab:acc_Kappa_vars_multgen}, and those for single-projects in Figure~\ref{fig:rq4_sp_AccKappaBoxPlots}.

\section{Threats to Validity}\label{sec:threads}
Our study is threatened by concerns of both internal and external validity. In the following, we outline these threats and how we alleviate or thwart them.

\MakeUppercase{\emph{\textbf{Internal Validity}}} We use a fundamental dataset in our study, that was previously published by~\citet{levin2017commits}. It contains more than $1\,150$ manually labeled commits. The authors of said dataset took numerous actions to mitigate threats to their labeling process, such as preventing class starvation (i.e., preventing that any of the classes is underrepresented), dropping commits with low confidence, and splitting the labeling work. They report having achieved an agreement level of $94.5$\% with an estimated asymptotic confidence interval of $[90.3\text{\%}, 98.7\text{\%}]$.

The data that we add ourselves was gathered systematically, using our tool suite \emph{\mbox{Git-Density}}~\cite{honel2019gitdensity}. It facilitates an industry-grade component for the detection of software clones and dead code, that was extensively applied and tested in that realm for more than ten years before this study. It continues to be under active development. Detecting whitespace and comments was reliably implemented using non-greedy regular expressions, inspecting line by line, then hunk by hunk. We have added an extensive suite of unit-tests to ensure that our tool behaves correctly. Therefore, we are confident in trusting that our mined size- and density-data is correct.

As other researchers have already pointed out~\cite{hattori2008nature,kirinuki2014hey}, a commit's nature may not always be pure (\emph{tangled} changes). We follow the classification into maintenance activities, as suggested by \citet{mockus2000identifying}. Those allow only to describe the nature for an entire commit. Often, changes in a commit are ambiguous, meaning that they could be seen as belonging to either of the three activities. Therefore, when labeling commits manually, some residual errors cannot be avoided. That error, to some degree, is also reflected in the models that we train.

Furthermore, having tangled changes is highly likely for merge-commits, as those are the result of merging the changes from two or more commits, as their name implies. We have rigorously excluded such commits in all of our analyses. Although, the models we built were likely to behave unpredictably for regular git-workflows that feature such commits. Merge-commits need to be investigated further before they can be included in our models.

We have also found strong multi-collinearities between the many features used for Research Question 1A. Predominantly the size features come in pairs of net/gross. Those have an expected strong positive correlation. Eleven attributes can be eliminated when using a cutoff of $0.9$ as the correlation coefficient. Models built using RFE perform only marginally worse with respect to accuracy and Kappa, for the benefit of reduced model complexity and a decrease in the variance of the regression coefficients.

\MakeUppercase{\emph{\textbf{External Validity}}} When training models using machine learning techniques, under- and over-fitting of such models is a concern. While the former means that a trained model cannot adequately represent the structure and patterns found in the data (and therefore performs poorly), the latter happens when the data is too small or the model too large, and the model contains more parameters than are justified by the data. Our attempt at mitigating either case was to apply RFE. In our results, we encountered both cases, under- and over-fitting. In RFE, the impact and importance of each variable for a model are measured. In our analyses, we ran the full extent of RFE, using between one and all available variables. Then, a many times repeated cross-validation was performed, always withholding a certain amount of data entirely from training. This resampling mechanism can assure external validity to a high degree. Under- and over-fitting was always observed in the context of models trained on individual projects. This is not surprising, as the amount of data available per project is significantly less. Also, due to that shortage, some features only exhibited a very low or no variance any longer, so that they had to be eliminated. The achieved results concerning individual projects should, therefore, be regarded as less generalizable compared to those for cross-projects. The results for single projects, however, demonstrate that models trained on them can approximate the nature of their commits with higher absolute accuracy and Kappa.

Furthermore, the eleven projects in the datasets were all open source projects. While others claim that the evolutionary patterns in such projects and closed software are the same~\cite{herraiz2006comparison}, we can neither support this claim nor generalize our results for closed software at this point. The eleven projects however represented a broader spectrum of software types: e.g., \emph{RxJava} and \emph{Spring-Framework} are libraries, \emph{Hadoop} is an enterprise distributed storage solution, \emph{Kotlin} is a programming language, \emph{IntelliJ Idea} is a fully-featured development environment for desktop, and \emph{ElasticSearch} and \emph{OrientDb} are enterprise-grade search- and database-engines, respectively. With a somewhat higher level of confidence, we expect high generalizability of our results for software that falls into this spectrum.

\section{Related Work} \label{sec:rel_work}
The related work can be subdivided into three categories. First, studies that present or examine various attempts to quantify changes in commits. Second, earlier similar studies to this one, that we in part reproduce and build on, or use directly. Third, studies that use the size of commits to solve a concrete problem.

\subsection{Quantification of Change}
Related studies found various ways to quantify the changes a commit induces. As an early method, function points were suggested by~\citet{albrecht1979measuring}, in \citeyear{albrecht1979measuring}. Rather than counting LOC, function points were meant to provide a way to quantify the size of a program as a functional size measurement. \citet{lin1988classifying}, develop an application that counts the number of changed, added, and deleted statements in COBOL applications, to quantify the size of a change. \citet{jackson1994semantic}, propose a difference algorithm that carves out semantic changes. \citet{fluri2007change}, apply tree-differencing to distill changes in the Abstract Syntax Trees (AST) between two versions of a software system, thus quantifying the syntactical changes in terms of statements and expressions. Several others, e.g., \cite{fischer2003populating,hindle2009automatic}, attempted to classify commits based on their keywords or other associated messages, such as those from bug-tracking systems. Characterizing commits by reverse-engineering the stereotype of affected methods was demonstrated by~\citet{dragan2011using}. While only indirectly measuring size, that approach provides a valuable insight into the nature of a commit and its effects on the system's architecture.

\subsection{Reproducibility and Relation to Previous Studies}
Most related work classifies the maintenance activities as \emph{adaptive}, \emph{corrective}, or \emph{perfective}, as proposed by \citet{swanson1976dimensions}, and further discussed by \citet{mockus2000identifying}. Others introduce additional or more distinctive categories. Our work extends and compares to the work of \citet{levin2017boosting}. Therefore, we adopt the three originally proposed categories (labels) without alterations as they do. We apply similar validation methods, focusing on prediction accuracy and \citeauthor{cohen1960coefficient}'s~ Kappa~\cite{cohen1960coefficient}, for measuring the agreement of our proposed models and the true labels.

\citet{levin2017boosting}, use a manually labeled dataset~\cite{levin2017commits}, containing 1\,151 commits as an underlying ground truth. They report a well-respectable classification model based on a hybrid classifier that exploits commit keywords and source code changes. The latter is obtained by the distiller from \citet{fluri2007change}. We are reproducing, reusing, and extending their work.

We are particularly interested in addressing questions that were answered previously without taking into account the net size of changes. This is of interest, as some studies report strong correlations between the size and the nature of a commit. \citet{hindle2008large} are looking in particular at large commits, where they derive the size of a commit by the number of files included in it. They find that large commits tend to be perfective, while small commits are often corrective. That work is especially relevant in conjunction with another study by \citet{herraiz2006comparison}, that found that it does not matter whether the size is defined via the number of files or the number of lines of code.

\subsection{Studies that rely on Size}
Furthermore, \citet{herraiz2006comparison} propose size classifications of commits based on the number of files changed in the commits. While we focus on extracting size features, we also possess the means to count the affected files. Further, using density, we can reduce those counts by files that were not affected in actuality. Using this new insight, we can put some of our results in relation to the findings of \citeauthor{herraiz2006comparison}. An additional study by~\citet{alali2008s} that attempted to categorize commits of nine open source projects by their size quantified using the number of files, the number of lines, and the number of hunks affected, reports a weak correlation between the first two, and a strong correlation between the last two measurements. Additionally, they confirm that comparatively small changes are afflicted with correcting bugs.

\citet{hattori2008nature}, build upon the work from \citeauthor{herraiz2006comparison} and use the number of affected files to determine the nature of commits. By attempting to specify size categories, such as \emph{tiny}, \emph{medium}, or \emph{large}, they report that the majority of tiny commits, i.e., affecting five files or less, are not related to adaptive development activities. Rather, those changes are perfective or corrective. Their study gives us further incentive to examine the relationship between notions of size based on either affected files or lines of code.


\citet{purushothaman2005toward}, address the problem of classifying small corrective changes by focusing on the properties of the changes rather than the properties of the code itself. They point out that change-size is a significant fault predictor. They raise awareness for risk assessment and risk management, as the risk associated with small changes tends to be small, too. With a more concise notion of net change size, or at least a better approximation, we are contrasting our findings against changes that were previously considered to be small.




\section{Conclusions}\label{sec:conclusions}

We have demonstrated that considering additional properties of the contents of a commit can significantly improve classification performance. We have shown that the \emph{density} of a commit is a significant predictor. By reproducing the results of others and putting our studies into relation to theirs, we have made our work comparable.

Qualitatively, the density of a commit allows a more fine-grained separation into maintenance activities than raw lines of code, as our results for RQ\,1 show. The significant deviations between net- and gross-size (RQ\,2) make clear, that the density has the potential to unveil changes that raw lines of code would hide, such as global renames or the incorporation of large portions of code. On the other hand, the density may hide changes not reflected in the change in functionality of the underlying source code, such as changes to the documentation. For models using density for the classification into the chosen three maintenance categories however, the density is suitable (RQ\,3). That also supports the results from the last research question (RQ\,4). If the density allows assigning a commit to some maintenance activity with great confidence, then the density of preceding commits carries weight that can be exploited for further improving commit classifiers.

Earlier studies suggested that measuring the size of a commit by counting either its affected files or lines of code are equal, but we find this not to be true. There are significant differences between those and more subtle differences when differentiating net- and gross-size. We are also observing a shift in maintenance activities for high-density commits away from \emph{corrective} and towards \emph{perfective}.

When studied thoroughly, it becomes apparent that there is a significant difference between gross- and net notions of the size of a commit, with the latter being a more important predictor of a commit's nature. Models based on size features consistently outperform the baseline set by us. Density is a sound classifier, especially when trained on individual projects. Cross-project, the achievable accuracy is much above the baseline as well. 

The third part of this study is an attempt to reproduce previous results and to extend them. We were able to comprehend previous work and extend it with our data and methods in a way that can boost classification by another $4$--$13$\% (towards $90$\%) with a \emph{significant} Kappa, using models that involve the density of commits.

Going beyond the attempts made by others, we exploit relational information from our extended dataset in the last part. In it, we demonstrate that preceding generations of commits that are solely exhibiting size features are boosting commit classification rates even further, up to $93$\% for single projects, with an almost perfect Kappa. This confirmed a conjecture of ours that maintenance activities, especially on designated branches, follow evolutionary patterns that are typically met during software development processes.

Our results demonstrate an improvement of the state of the art in automated commit classification. Beyond that, we contribute the following:

\textbf{Git-Density}
is an open-source suite of tools for analyzing git-repositories~\cite{honel2019gitdensity}. It was initially built for extracting size- and especially density properties of commits' associated source code but has been expanded ever since.

\textbf{Extended Dataset}
The dataset used for all of our experiments is publicly available~\cite{honel2019commits} under the terms of open access.
Refer to subsection \ref{ssec_dataset} for detailed descriptions of contained features.

\textbf{R Experiments} 
The experiments were conducted using the R statistical environment~\cite{renv2017}. We have performed extensive analyses on the published data and to strengthen collaborative scientific work and to aid the reproducibility of our results, we share the source code for all experiments on GitHub\footnote{https://github.com/MrShoenel/density-paper-2019-R-experiments}.

\section{Future Work}\label{sec:future}
We find that the size of a commit, while a significant predictor for maintenance activities, is a computationally cheap and convenient measure to use. We plan to package our classifier into a tool that can be used to automatically classify commits and use it to perform a field study. We aim to apply the classifier across a number of open source projects and use the classification information to support tasks, such as fault prediction, or to characterize the evolution process and aspects of it automatically.

Dimensionality reduction techniques could help to reduce the number of attributes, since we have few samples. Our attempts to visually analyze the data using the t-SNE~\cite{maaten2008visualizing} algorithm were not fruitful. During the experiments, we also attempted to reduce the number of dimensions using a Principal Component Analysis (PCA)~\cite{pearson1901liii}. However, this did not yield significant improvements.

We have further identified that structural and primarily relational properties of commits had not previously been considered, to the best of our knowledge. Commits are assigned to branches in a source tree, they have predecessors and successors, and there are special kinds of commits, such as merge-commits. Often, branches serve a single purpose, and adjacent commits may share a common nature, or their nature follows a logical pattern. Our tool \emph{\mbox{Git-Density}}~\cite{honel2019gitdensity} was extracting such properties, so we used them in research question four.

However, now that we found previous generations of commits to be useful, more potential extensions open up. The first that comes to mind are Hidden Markov Models~(HMM, \citet{baum1966statistical}). Such models find the most likely path of hidden states (here: commits' labels) through a series of observable events (commits' features). But building said models would require us to label adjacent commits from our extended dataset manually. In their simplest form, HMMs are univariate models that would have a questionable application, given the vast amounts of features we were evaluating. Also, numeric data need to be discretized into events, which need to have a probability of occurrence assigned.

Another potential approach for taking multiple generations and multiple features into account would be to apply Bayes' theorem. It is built around conditional probabilities and can be extended efficiently to operate on occurrences of more than just one event. Such an example, given the two conditional events $\{B,C\}$ is given in Equation~\ref{eqn:multi_bayes}, the example may be extended to accommodate an arbitrary number of events. The advantage of using a Bayes approach lies in its simplicity and the low effort required to set it up. It would, however, require discretized events from numeric data, too.
\begin{equation}\label{eqn:multi_bayes}
P(A|B,C) = \frac{P(B|C,A) \times P(C|A) \times P(A)}{P(B|C) \times P(C)}
\end{equation}
One more substantial extension of our approach could be the application of multivariate HMMs. While these models tend to be more complex, they support a probability distribution for each feature of the current observation. We conjecture that such models would likely deliver high classification rates, or that they could be used in an ensemble- or meta-learner, to further stabilize prediction accuracy.

Lastly, we have extracted the dwell times between adjacent commits in our extended dataset, but have not yet made use of them. Such information can be exploited in Hidden semi-Markov Models (HsMM,~\citet{yu2010hidden}). These models allow for a separate probability distribution of dwell times in each state, and we surmise that this feature carries some weight.

\section*{Acknowledgments}
We would like to thank the anonymous reviewers for their invaluable comments, which helped us to further improve this work. We would also like to thank the Linnaeus University Centre for Data Intensive Sciences (DiSA) and Applications and the Swedish Research School of Management and IT (MIT).






\bibliographystyle{elsarticle-num-names}
\bibliography{main}


\end{document}